%% file: phases.tex
\PassOptionsToPackage{unicode}{hyperref}
\PassOptionsToPackage{hyphens}{url}
\documentclass[
  astrosymb, twocolumn, twocolappendix, tighten]{aastex631}
\usepackage{xcolor}
\usepackage{amsmath,amssymb}

\usepackage{iftex}
\ifPDFTeX
  \usepackage[T1]{fontenc}
  \usepackage[utf8]{inputenc}
  \usepackage{textcomp} 
\else 
  \usepackage{unicode-math} 
  \defaultfontfeatures{Scale=MatchLowercase}
  \defaultfontfeatures[\rmfamily]{Ligatures=TeX,Scale=1}
\fi
\usepackage{lmodern}
\ifPDFTeX\else
\fi
\IfFileExists{upquote.sty}{\usepackage{upquote}}{}
\IfFileExists{microtype.sty}{
  \usepackage[]{microtype}
  \UseMicrotypeSet[protrusion]{basicmath} 
}{}
\makeatletter
\@ifundefined{KOMAClassName}{
  \IfFileExists{parskip.sty}{%
    \usepackage{parskip}
  }{
    \setlength{\parindent}{0pt}
    \setlength{\parskip}{6pt plus 2pt minus 1pt}}
}{
  \KOMAoptions{parskip=half}}
\makeatother
\usepackage{graphicx}
\makeatletter
\newsavebox\pandoc@box
\newcommand*\pandocbounded[1]{
  \sbox\pandoc@box{#1}%
  \Gscale@div\@tempa{\textheight}{\dimexpr\ht\pandoc@box+\dp\pandoc@box\relax}%
  \Gscale@div\@tempb{\linewidth}{\wd\pandoc@box}%
  \ifdim\@tempb\p@<\@tempa\p@\let\@tempa\@tempb\fi
  \ifdim\@tempa\p@<\p@\scalebox{\@tempa}{\usebox\pandoc@box}%
  \else\usebox{\pandoc@box}%
  \fi%
}
\def\fps@figure{htbp}
\makeatother
\providecommand{\tightlist}{%
  \setlength{\itemsep}{0pt}\setlength{\parskip}{0pt}}
\usepackage[]{natbib}
\usepackage{mathptmx,txfonts,tikz,bm}
\usepackage[capitalize]{cleveref}

\newcommand{\annotate}[2]{\begin{tikzpicture}
    \node[anchor=south west,inner sep=0,align=center] (image) at (0,0) {
    #1
    };
    \begin{scope}[x={(image.south east)},y={(image.north west)}]
    #2
    \end{scope}
\end{tikzpicture}}

\renewcommand{\edit}[2]{{\ifnum#1 <  1 %
#2%
\else%
\textbf{#2}%
\fi}}

\input{macros.tex}

\defcitealias{roxburgh_asymptotic_1994}{RV94}
\newcommand\rv{{\citetalias{roxburgh_asymptotic_1994}}}
\usepackage[]{natbib}
\usepackage{bookmark}
\IfFileExists{xurl.sty}{\usepackage{xurl}}{} 
\received{August 19, 2024}
\revised{December 19, 2024}
\accepted{January 9, 2025}
\submitjournal{The Astrophysical Journal}

\let\[\relax \let\]\relax 
\DeclareRobustCommand{\[}{\begin{equation}}
\DeclareRobustCommand{\]}{\end{equation}}

\begin{document}

\title{Resolving an Asteroseismic Catastrophe: Structural Diagnostics from p-mode Phase Functions off the Main Sequence}

\correspondingauthor{Joel Ong}
\email{joelong@hawaii.edu}

\author[0000-0001-7664-648X]{J. M. Joel Ong \chinesename}
\altaffiliation{NASA Hubble Fellow}
\affiliation{Institute for Astronomy, University of Hawaiʻi, 2680 Woodlawn Drive, Honolulu, HI 96822, USA}

\author[0000-0001-8722-1436]{Christopher J. Lindsay}
\affiliation{Department of Astronomy, Yale University, P.O. Box 208101, New Haven, CT 06520, USA}

\author[0000-0001-9632-2706]{Claudia Reyes}
\affiliation{School of Physics, University of New South Wales, Sydney, NSW 2052, Australia}

\author[0000-0002-4879-3519]{Dennis Stello}
\affiliation{School of Physics, University of New South Wales, Sydney, NSW 2052, Australia}
\affiliation{Sydney Institute for Astronomy, School of Physics, A28, The University of Sydney, NSW 2006, Australia}
\affiliation{ARC Centre of Excellence for All Sky Astrophysics in 3 Dimensions (ASTRO 3D), Australia}
\affiliation{Stellar Astrophysics Centre, Aarhus University, Ny Munkegade 120, DK-8000 Aarhus C, Denmark}

\author[0000-0002-7403-2764]{Ian W. Roxburgh}
\affiliation{Astronomy Unit, Queen Mary University of London, Mile End Road, London E1 4NS, UK}
\affiliation{School of Physics and Astronomy, University of Birmingham, Edgbaston, Birmingham B15 2TT, UK}

\shortauthors{Ong, Lindsay, Reyes et al.}
\shorttitle{Post-MS Phase Functions}
\begin{abstract}
On the main sequence, the asteroseismic small frequency separation \(\delta\nu_{02}\) between radial and quadrupole p-modes is customarily interpreted to be a direct diagnostic of internal structure. Such an interpretation is based on a well-known integral estimator relating \(\delta\nu_{02}\) to a radially-averaged sound-speed gradient. However, this estimator fails, catastrophically, when evaluated on structural models of red giants: their small separations must therefore be interpreted differently. We derive a single expression which both reduces to the classical estimator when applied to main-sequence stellar models, yet reproduces the qualitative features of the small separation for stellar models of very evolved red giants. This expression indicates that the small separations of red giants scale primarily with their global seismic properties as \(\delta\nu_{02} \propto \Dnu^2/\numax\), rather than being in any way sensitive to their internal structure. Departures from this asymptotic behaviour, during the transition from the main-sequence to red giant regimes, have been recently reported in open-cluster Christensen-Dalsgaard (C-D) diagrams from K2 mission data. Investigating them in detail, we demonstrate that they occur when the convective envelope boundary passes a specific acoustic distance --- roughly a third of a wavelength at \(\numax\) --- from the centre of the star, at which point radial modes become maximally sensitive to the position of the boundary. The shape of the corresponding features on \(\epsilon_p\) and C-D (or \(r_{02}\)) diagrams may be useful in constraining the nature of convective boundary mixing, in the context of undershooting beneath a convective envelope.
\end{abstract}
\keywords{Asteroseismology (73), Red giant stars (1372), Subgiant stars (1646), Stellar convective zones (301), Theoretical techniques (2093)}

\def\sectionautorefname{Section}
\def\subsectionautorefname{Section}
\def\subsubsectionautorefname{Section}

\section{Introduction and Problem Statement}\label{introduction-and-problem-statement}

Asteroseismology is our only direct means of observationally inspecting the properties of stellar interiors. Seismology from NASA's \emph{Kepler} mission has revolutionised our understanding of stellar structure, evolution, rotation, activity, and demographics, all over the Hertzsprung-Russell diagram \citep[for a review, see][]{aerts_probing_2021}. Many of these breakthroughs --- both in main-sequence and red giant stars --- have been made using solar-like stochastically-excited pressure-wave pulsations (or p-modes). In Sun-like stars, the frequencies of these p-mode oscillations, in various families of overtones grouped by the latitudinal degree \(\ell\) of their horizontal shapes, are approximately separated by a uniform overtone spacing \(\Delta\nu\). One might parameterise their frequencies \(\nu_{n\ell}\) as, using the notation of \citet{scherrer_detection_1983},
\[
\nu_{n\ell}\sim\Dnu\left(n + {\ell \over 2} + \epsilon_\mathrm{p}(\nu)\right) - {\ell(\ell + 1)}{\delta\nu_{02} \over 6},\label{eq:solar}
\]
where \(n\) is an integer radial order, \(\epsilon_\mathrm{p}\) is a phase offset function that varies smoothly with frequency \(\nu\), and \(\delta\nu_{02}\) is the ``small separation'' between quadrupole (\(\ell = 2\)) and radial (\(\ell = 0\)) p-modes,
\[
  \delta\nu_{02}(n) = \nu_{n, \ell=0} - \nu_{n-1, \ell=2}.\label{eq:smallsep}
\]
Low-luminosity red giants are also solar-like oscillators, in that their pulsations are also stochastically excited by surface convection. The frequencies of their non-radial modes exhibit qualitatively different features compared to main-sequence Sun-like stars --- namely that of ``avoided crossings'' between these p-modes and an additional set of g-modes trapped within their radiative cores \citep{aizenman_avoided_1977, shibahashi_modal_1979, bedding_gravity_2011}. Nonetheless, the most visible of these gravitacoustic ``mixed'' modes emerge at frequencies that are closest to those of notional pure p-modes \citep{bedding_solarlike_2010, mosser_universal_2011, stello_nonradial_2014}, as described using \cref{eq:solar}. Thus these modes may also be associated with small separations through \cref{eq:smallsep}.

Much of the modern theory of these solar-like oscillators is inherited from helioseismology. In this tradition, the ``small separation'' is interpreted as a direct diagnostic of a star's interior structure. This diagnostic property arises from the fact that the associated ``small separation ratio'',
\[
   r_{02}(n) =  {\nu_{n, \ell=0} - \nu_{n-1, \ell=2} \over \nu_{n, \ell=1} - \nu_{n-1, \ell=1}}\sim {\delta\nu_{02} \over \Dnu},\label{eq:ratio}
\]
is, at least in the context of Sun-like stars, insensitive to a star's near-surface layers \citep[e.g.][]{roxburgh_ratio_2003, roxburgh_ratio_2005, otifloranes_use_2005}. Hence, changes in \(r_{02}\), and therefore \(\delta\nu_{02}\), are interpreted as reflecting the interior structure of a star \citep[e.g.][]{miglio_constraining_2005, nsamba_modelling_2018, valle_relevance_2020}. For instance, they probe changes to the chemical stratification of the near-core layers, as modified by hydrogen burning, which arise over the course of main-sequence evolution. As a result, \(\delta\nu_{02}\) is commonly used as an asteroseismic diagnostic of not only structure, but also main-sequence age \citep[such as on a Christensen-Dalsgaard, or C-D, diagram, as explored in][]{white_diagrams_2011}.

These properties arise from asymptotic analysis of the wave equation describing stellar oscillations. For p-modes, one may rescale the pulsation eigenfunctions as \(\psi = \xi_r\sqrt{r^2 \rho c_s}\), where \(r\) is the radial coordinate, \(\rho\) is the equilibrium density profile, \(c_s\) is the sound speed, and \(\xi_r\) is the Lagrangian displacement eigenfunction of the normal mode in the radial direction. In asymptotic analysis, this rescaled eigenfunction \(\psi\) approximately satisfies \citep[e.g.][]{calogero_novel_1963, babikov_method_1976}
\[
\begin{aligned}
  \psi &\sim A(t)s_\ell\left(\omega t + \delta_\ell(\omega, t)\right) & \text{(inner)}\\
  &\sim A(t) \sin\left(\omega(t-T) + \alpha_\ell(\omega, t)\right); & \text{(outer)}\label{eq:psi}
\end{aligned}
\]
with the two expressions given being the ``inner'' and ``outer'' solutions. Here \(s_\ell(x) = \sqrt{\pi x \over 2} J_{\ell + {1\over2}}(x)\) is the Riccati-Bessel function of the first kind, \(A(t)\) is a slowly varying amplitude function, \(\omega = 2\pi\nu\) is the angular frequency, \(t\) or rather \(t(r) = \int_0^r (1/c_s) \mathrm d r\) is the acoustic radial coordinate, \(T = t(R)\) is the acoustic radius, and \(\delta_\ell\) and \(\alpha_\ell\) are the inner and outer partial phase functions, respectively, associated with the degree \(\ell\). In this work, we also use the property that, far from the center and surface of the star, these phase functions are insensitive to the acoustic radial coordinate. As such, we treat their values there as functions of only the mode frequency --- we may therefore write them as \(\delta_\ell(\omega)\) and \(\alpha_\ell(\omega)\) with no radial dependence \citep[see][ for an overview of their other properties]{roxburgh_ratio_2003, roxburgh_ratio_2005, lindsay_nearcore_2023}. In terms of these quantities, \citet{roxburgh_ratio_2003} show that the separation ratio \(r_{02}\), treated now as a continuous function of frequency, may be expressed compactly as
\[
  r_{02}(\omega) \sim {1 \over \pi}\left(\delta_2(\omega) - \delta_0(\omega)\right),\label{eq:phasediff}
\]
with no dependence on the structure of the outer layers of the star. In turn, these phase functions are related to the normal-mode frequencies by demanding that the inner and outer expressions for \(\psi\) given in \cref{eq:psi} agree up to sign (meaning they satisfy \(\psi_\text{in} = \pm \psi_\text{out}\)) far from the center and surface, yielding an eigenvalue condition of the form
\[
  \sin\left[\omega T - \alpha_\ell(\omega) + \delta_\ell(\omega) - {\ell \over 2}\pi\right] = 0.\label{eq:eig}
\]
Eigenvalues \(\omega_{n\ell m}\) are obtained where the argument of the sine function is an integer multiple of \(\pi\), yielding the observational parameterisation of \cref{eq:solar}.

\begin{figure}
\centering
\annotate{\includegraphics[width=0.9\linewidth,height=\textheight,keepaspectratio]{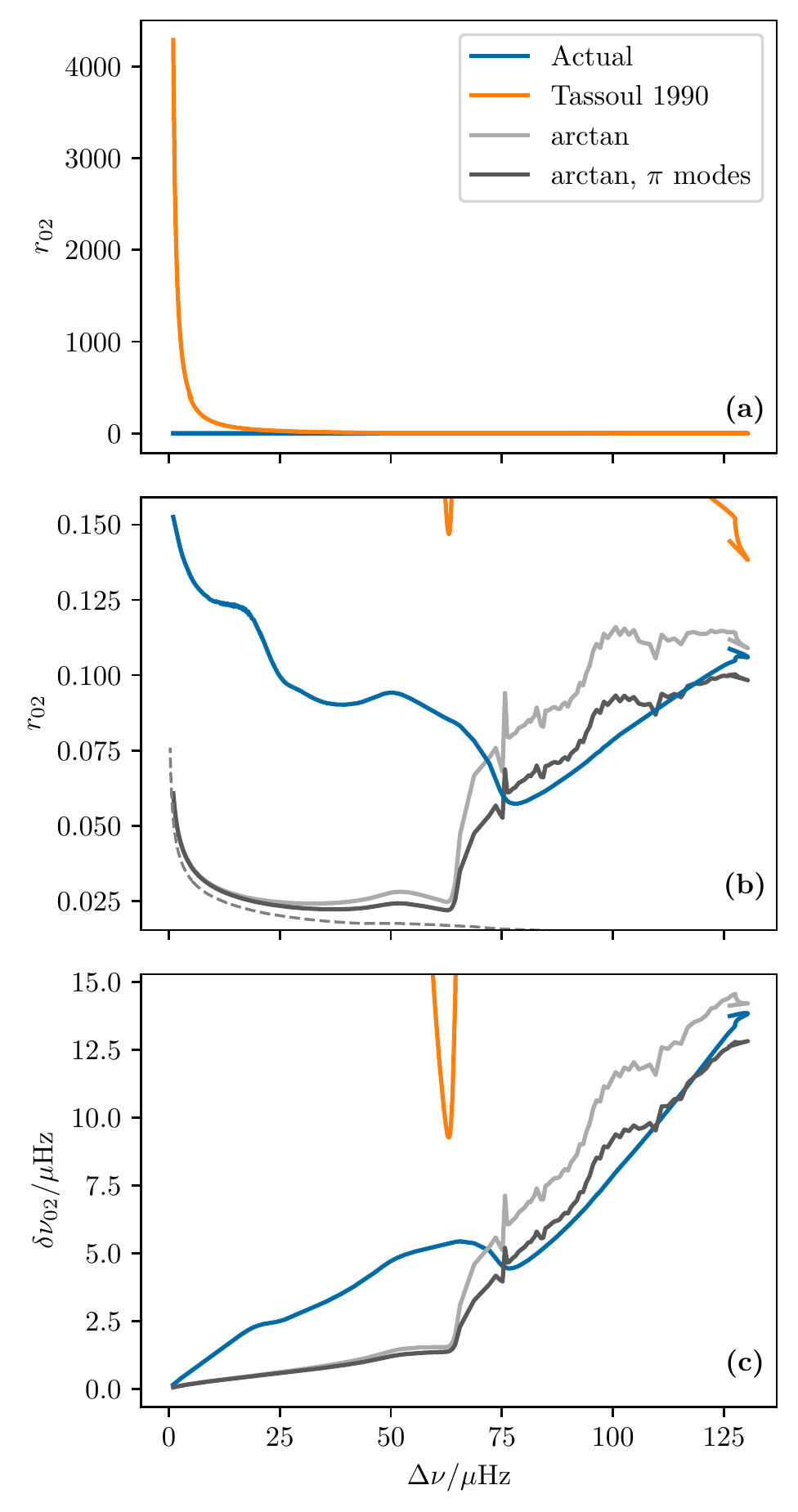}}{\draw[red, thick](.26,.52) rectangle ++(.12,.1); \draw[red, thick](.26,.1) rectangle ++(.12,.05);}
\caption{\footnotesize Separation ratios and small separations computed with different methods, along a \mesa~evolutionary track of solar composition, solar-calibrated mixing length, and with \(M = 1.2 M_\odot\). \textbf{(a)}: Separation ratios \(r_{02}\) along the evolutionary track. The blue curve shows values computed from the \(\ell = 0,2\) mode frequencies returned from the \gyre~stellar oscillation code through \cref{eq:smallsep}. The orange curve shows values obtained from using the conventional asymptotic estimator, \cref{eq:tassoul} --- which yields unphysically large values, indicating that approximations used to derive it have failed catastrophically. The other gray curves show estimators derived in this work, which do not exhibit the same divergence as the orange curve. \textbf{(b)}: The same quantities as (a), with the vertical axis rescaled to better show the other estimators we derive in this work. The light gray curve shows values obtained when a faulty small-angle approximation used in deriving \cref{eq:tassoul} is no longer assumed. The dark gray curve shows values obtained when both \(\delta_0\) and \(\delta_2\) are computed under the \(\pi\)-mode isolation scheme of \citet{ong_semianalytic_2020} (\(\alpha_\pi = 0\)), rather than for the full system of equations (\(\alpha_\pi = 1\)). The dashed curve shows the limiting asymptotic value \(3/\pi\omega T\). \textbf{(c)}: A C-D diagram, showing the small frequency separations \(\delta\nu_{02} = r_{02} \times \Dnu\) from the same models and methods. The blue curves depart from these asymptotic estimators, and our analysis indicates that they are associated with features of the Brunt-Väisälä frequency. The feature at \(\Dnu \sim 20\ \mu\)Hz (highlighted in red) is the Knee feature reported in \citet{reyes_in_review}. \label{fig:catastrophe}}
\end{figure}

Existing analytic expressions relating these small separations (or separation ratios) to stellar structure have arisen from studies restricted in scope to the main sequence \citep[as was the case for][]{tassoul_asymptotic_1994, roxburgh_asymptotic_1994, roxburgh_ratio_2003, roxburgh_ratio_2005, otifloranes_use_2005}. Applied to red giants, however, these expressions fail catastrophically (as we will illustrate in \cref{fig:catastrophe}a). Hence, they must be interpreted differently for these evolved stars. No alternative interpretation has so far been proposed. Moreover, the theoretical studies in which these expressions were derived all predate the observational discovery of nonradial gravitoacoustic ``mixed'' modes in such evolved solar-like oscillators. Therefore, they also predate the subsequently developed mathematical machinery required to decompose the pulsation equations governing these mixed modes into pure p- and pure g-mode subsystems \citep{ong_semianalytic_2020}. These techniques now enable us to examine red-giant small separations using the same asymptotic techniques as for main-sequence stars. At the same time, the observational seismic characterisation of open clusters has also only recently revealed astrophysically significant features in their C-D diagrams \citep{reyes_in_review}, which thus also only now demand astrophysical interpretation in terms of interior structure.

Accordingly, in this work we will make use of these recent theoretical developments to derive new analytical asymptotic expressions relating the inner phase functions \(\delta_\ell\) to the interior structure of red giants, which reduce to known ones when applied to main-sequence stars. These new analytic expressions are required for us to interpret a recently-reported observational feature in the C-D diagrams of stellar clusters, and allow us to understand that it emerges only in stars whose convective-envelope boundaries are located close to a specific distance from the centre of the star. We finally discuss how this feature, which emerges purely in quantities derived from p-mode frequencies, may be used as a probe of convective boundary mixing, in conjunction with other observational features on the red giant branch constructed from \numax, g-modes, or other spectroscopic constraints.

\section{Resolving an asteroseismic catastrophe}\label{resolving-an-asteroseismic-catastrophe}

\citet{tassoul_asymptotic_1994} relates the p-mode small separation, \cref{eq:smallsep}, to the stellar structure, through a second-order asymptotic expression of the form
\[
  r_{\ell, \ell+2} = {\delta\nu_{\ell, \ell+2} \over \Dnu} \sim  -{2\ell + 3 \over \pi^2 \nu}\left[\int_0^R{1\over r}{\mathrm d c_s \over \mathrm d r}\mathrm d r - {c_s(R)\over R}\right]+ \mathcal{O}\left(1\over\nu^2\right), \label{eq:tassoul}
\]
In practice, \(\nu\) in this expression may be set to \(\nu_\mathrm{max}\) when averaging over many radial orders \citep{roxburgh_ratio_2005, otifloranes_use_2005}. Here \(\mathcal{O}\) is Landau's symbol, denoting the order of the asymptotic approximation. \citet[hereafter \rv]{roxburgh_asymptotic_1994} further expand this to fourth order in \(1/\nu\). Generally speaking, the coefficients of all terms in such asymptotic expansions are integrals of various functions over the entire stellar structure. The ability of small separations and separation ratios to probe stellar interiors is often attributed to expressions of the kind given in \cref{eq:tassoul}.

\subsection{The Catastrophe}\label{the-catastrophe}

This entire class of integral estimators ceases to hold for red giants. We illustrate this in \cref{fig:catastrophe}a, where we plot the values of \cref{eq:tassoul} as a function of \(\Dnu\) over an illustrative track of \mesa~evolutionary models, details of which construction we describe in the next section. \Cref{eq:tassoul} can be seen to take extremely large values, shown with the orange curve. Values this large are unphysical, as the mode frequencies of observed red giants continue to be well-described as possessing small separations even close to the tip of the red giant branch \citep[e.g.][]{bedding_solarlike_2010, mosser_universal_2011, stello_nonradial_2014, yu_luminous_2020}. While we have shown only one evolutionary track in \cref{fig:catastrophe} for clarity, this behaviour continues to emerge when other stellar properties are varied (such as mass, composition, and mixing length parameter).

This ``asteroseismic catastrophe'' is not caused by any singularities in the underlying pulsation equations, nor physical inconsistencies in the models themselves. Numerical solutions to the pulsation equations on the same stellar models \citep[e.g.~using the \gyre~pulsation code,][shown in blue in \cref{fig:catastrophe}]{townsend_gyre_2013} remain well-behaved, and return mode frequencies that are well-described as possessing small separation ratios much smaller than unity. This remains true even after the emergence of gravitoacoustic mode mixing, whereupon small separations are observationally determined with reference to the most p-dominated quadrupole mixed modes. In such cases, \citet{ong_semianalytic_2020} demonstrate that one may directly calculate the pure p-modes underlying the p-dominated mixed modes, by suppressing an individual term in the pulsation equations --- a single occurrence of the Brunt-Väisälä frequency \(N\) --- in the interior of the star. The asymptotic analysis of \citet{tassoul_asymptotic_1994} and \rv, applied to this modified ``\(\pi\)-mode'' system \citep[in the sense of][]{aizenman_avoided_1977}, returns exactly the same expression for the small separation as \cref{eq:tassoul}: after all, it can be seen not to depend on \(N\). Finally, even though calculations using p-dominated mixed modes return significant amounts of numerical jitter \citep[e.g.][]{white_diagrams_2011}, they still result in small separation ratios smaller than unity. As such, the behaviour of the orange curve in \cref{fig:catastrophe}a cannot be attributed to mode mixing, either. Thus, by elimination, it must instead result from the catastrophic failure of some other approximation involved in the asymptotic analysis required to derive \cref{eq:tassoul} in the first place.

The specific form of \cref{eq:tassoul} also suggests that this behaviour is a generic feature of red giant models (in particular, of their posessing small, dense, radiative cores), rather than a particular feature of how we have generated these models, or computed these estimators. This is because the cores of red giants, which are isothermal and therefore radiatively stratified after core hydrogen exhaustion, shrink concurrently with the expansion of their outer convective envelopes by the mirror principle throughout their first ascent up the red giant branch \citep[e.g.][]{bertolami_giants_2022, ou_stars_2024}. As a result, the radiative cores of red giants are very compact, while their envelopes are very diffuse. This sets up very large density and therefore sound-speed gradients. The sound-speed gradient in turn enters into the integrand of \cref{eq:tassoul}, and so its estimates of the small separation in red-giant stellar models thus take on correspondingly large values, which become unphysically large as they ascend the red giant branch.

Irrespective of the reason, it is clear that \cref{eq:tassoul} does not hold in evolved stars. The inclusion of higher-order terms in the asymptotic expansion from which it is taken --- such as those of \rv --- does not change the fact that this leading-order term already diverges. Correspondingly, then, we ought not to use it to interpret observations of small separations, and separation ratios. This begs the question, however, of what meaning we ought to attach to these observational quantities off the main sequence.

\subsection{Asymptotic Analysis}\label{asymptotic-analysis}

Our subsequent discussion will build on the procedure for asymptotic analysis laid out in \rv, so we will now briefly summarise it. Linear adiabatic self-gravitating pulsations have normal modes whose Lagrangian displacements \(\xi_r\), and Eulerian pressure and gravitational-potential perturbations \(P'\) and \(\Phi'\), satisfy a linear system of ordinary differential equations after separation of variables:
\[
\begin{aligned}{1 \over r^2}{\mathrm d \over \mathrm d r} (r^2 \xi_r) - {g \over c_s^2}\xi_r + \left(1 - {\ell(\ell+1) c_s^2 \over r^2 \omega^2}\right) {P' \over \rho c_s^2} &= {\ell(\ell+1) \over r^2 \omega^2}\Phi',\\{1 \over \rho} {\mathrm d P' \over \mathrm d r} + {g \over \rho c_s^2}P' + (N^2-\omega^2)\xi_r &= -{\mathrm d \Phi' \over \mathrm d r},\\{1 \over r^2}{\mathrm d \over \mathrm d r}\left(r^2 {\mathrm d \Phi'\over \mathrm d r}\right) - {\ell(\ell+1) \over r^2} \Phi' = 4\pi G\rho\left({P' \over \rho c_s^2} + {N^2 \over g}\xi_r\right).\end{aligned}\label{eq:pulsationode}
\]
\rv~simplify this through a change of dynamical variables to the quantities (eqs. 3-6 of \rv)
\[
\begin{aligned}
\xi &= {r^2 \over h_1} \xi_r \\
\eta &= {1\over \rho h_2} P'\\
p &= -\Phi' \\
s &= {1 \over \ell(\ell + 1)}\left(r^2 {\mathrm d p \over \mathrm d r} - 4 \pi G\rho h_1 \xi\right),
\end{aligned}
\]
written in terms of integrating factors
\[
h_1 = \exp\left[\int^r {g \over c_s^2}\ \mathrm d r\right];\ h_2 = \exp\left[\int^r {N^2 \over g} \ \mathrm d r\right];\ h = {h_2 \over h_1},
\]
where \(g\) is the local gravitational field strength. Doing so casts \cref{eq:pulsationode} as a second-order differential equation in only two of these quantities, written in the matrix form
\[
  \left({\mathrm{d}^2\over\mathrm{d}r^2} + \mathbf{C}{\mathrm{d}\over\mathrm{d}r} - \omega^2 \mathbf{D}\right)\begin{bmatrix}\eta\\p\end{bmatrix} = 0.
\]
The matrices \(\mathbf{C}\) and \(\mathbf{D}\) are fairly cumbersome, so we will not reproduce the full expressions for them here, but it suffices to note that

\begin{enumerate}
\def\labelenumi{(\arabic{enumi})}
\tightlist
\item
  they depend on both the equilibrium structure of the star, and on the mode frequency \(\omega\), but not the normal modes themselves, as well as that
\item
  every frequency-dependent entry of \(\mathbf{C}\) is proportional to \(\left(1 - N^2/\omega^2\right)^{-1}\), while those of \(\mathbf{D}\) contain frequency dependence only through terms proportional to \(1/\omega^2\) and \(1/\omega^4\).
\end{enumerate}

Thus, writing both \(\eta\) and \(P\) at degree \(\ell\) as linear combinations of Bessel functions \(J_{\ell + {1/2}}(\omega t)\) and their derivatives \(J'_{\ell + {1/2}}(\omega t)\), \rv~found that the coefficients of these linear combinations admit an asymptotic expansion in powers of \(1/\omega\):
\[
  \begin{bmatrix}\eta\\p\end{bmatrix} \sim \left(\mathbf{Y}_0 + {1\over\omega}\mathbf{Y}_1 + {1\over\omega^2}\mathbf{Y}_2 + \ldots \right)\begin{bmatrix}J_{\ell + {1\over2}}(\omega t)\\J'_{\ell + {1\over2}}(\omega t)\end{bmatrix}. \label{eq:yexpansion}
\]
Expressions for the coefficients \(y_{i,jk}\) of this expansion are derived by similarly expanding the matrices \(\mathbf{C}\) and \(\mathbf{D}\) in powers of \(1/\omega\), assuming from the outset that the angular frequency \(\omega\) is much larger than the Brunt-Väisälä frequency \(N\), and noting further that \(J_{\ell + 1/2}\) and its derivative satisfy
\[
  {\mathrm d \over \mathrm d t}\begin{bmatrix}J_{\ell + {1\over2}}(\omega t)\\J'_{\ell + {1\over2}}(\omega t)\end{bmatrix} = \begin{bmatrix}0 & \omega \\ - \omega + {\left(\ell + {1\over 2}\right)^2\over \omega t^2}  & -{1\over t}\end{bmatrix}\begin{bmatrix}J_{\ell + {1\over2}}(\omega t)\\J'_{\ell + {1\over2}}(\omega t)\end{bmatrix}.
\]
In this manner, one may express \(\eta\) and \(p\) (eqs. 49 and 50 of \rv) in terms of several of the \(y_{i,jk}\) (eqs. 32, 34, 41 of \rv), again formulated in terms of inner and outer solutions. Having done so, \rv~then construct an eigenvalue equation of the form of \cref{eq:eig} by noting that, since the Cowling approximation holds well in the outer layers, the outer solution may be expressed entirely in terms of a single dynamical variable \(\zeta\), with (eqs. 55, 57 of \rv)
\[
  \zeta \equiv \sqrt{c_s\over hr^2}\left(h r^2\over \omega^2\right)\left(1 - {N^2\over\omega^2}\right)^{-1}{\mathrm d \eta \over \mathrm d r} \to \xi\sqrt{c_s\over hr^2} = \psi \text{ as } r \to R.
\]
One then inserts Hankel's expansion of Bessel functions and their derivatives at large argument \citep[\(x \gg \ell(\ell+1)\):][]{abramowitz_stegun_1972},
\[
\begin{aligned}
  J_{\ell + {1\over2}}(x) &\sim \sqrt{2\over \pi x} \left[\sin\left(x - {\ell \pi \over 2}\right) + {\ell(\ell+1)\over 2x}\cos\left(x - {\ell \pi \over 2}\right) \right] + \mathcal{O}\left(1\over x^{5/2}\right);\\
  J'_{\ell + {1\over2}}(x) &\sim \sqrt{2\over \pi x} \left[\cos\left(x - {\ell \pi \over 2}\right) - {\ell(\ell+1) + 1\over 2x}\sin\left(x - {\ell \pi \over 2}\right) \right] + \mathcal{O}\left(1\over x^{5/2}\right),\label{eq:hankel}
\end{aligned}
\]
into their inner asymptotic expansion for \(\zeta\); their eq. 58 then follows from rearranging this into the form of a single sinusoid whose argument and overall amplitude are separately expanded asymptotically. Focusing on the argument of this sinusoid specifically, it is expanded as
\[
\begin{aligned}
  \zeta &\propto \cos \left[\omega t - {\ell \pi \over 2} + {1\over\omega}\left({y_{1,12}(t)\over y_{0,11}(t)} - {1 \over 2}{\mathrm d \over \mathrm d t}\log\left[c_s \over h r^2\right] + {\ell (\ell + 1)\over 2 t}\right) + \mathcal{O}\left(1\over\omega^2\right)\right]\\
  &\equiv \cos\left[\omega t - {\ell\pi\over2} + {1\over\omega}\left[A_0(t) + \ell(\ell+1)A_\ell(t)\right] + \mathcal{O}\left(1\over\omega^2\right) \right] \\
  &= \sin\left[\omega t - {\ell \pi \over 2} + \delta_\ell(\omega, t) \right].\label{eq:roxburgh58}
\end{aligned}
\]
This rather important expression is then used to derive an eigenvalue condition of the form of \cref{eq:eig} in the usual fashion, by matching the inner and outer solutions up to sign (\rv, eq. 80). In turn, this gives an expression for the small separation when finite differences with respect to \(\ell\) are taken (\rv, eq. 87). The estimator of \citet{tassoul_asymptotic_1994}, which results from truncating this analysis to leading order in \(1/\omega\), is thus only proportional to \(A_\ell(T)\), with no dependence on \(A_0(T)\).

\subsection{Resolving the Catastrophe}\label{resolving-the-catastrophe}

We now present the derivation of modified expressions, which reduce to \cref{eq:tassoul} for main-sequence stars, but yet also remain valid for application to red-giant stellar structures. Specifically, we observe that in deriving their inner expression for \(\zeta\), via \cref{eq:roxburgh58}, \rv~have rearranged the linear combination of several sines and cosines with the same argument, as suggested by the form of \cref{eq:hankel}, into a single harmonic function using the phasor addition identity, where in particular
\[
  a\sin \theta + b \cos \theta = \sqrt{a^2 + b^2}\sin\left(\theta + \arctan\left[b\over a\right]\right).
\]
Applied to their inner expression for \(\zeta\), one obtains for the inner solution that, truncated to leading order in \(1/\omega\),
\[
\begin{aligned}
  \zeta &\propto \left[{1\over\omega}\left(-{y_{1,12}(t)\over y_{0,11}(t)} + {1 \over 2}{\mathrm d \over \mathrm d t}\log\left[c_s \over h r^2\right] + {1 \over 2t}\right) + \mathcal{O}\left(1\over\omega^3\right)\right]J_{\ell + {1\over2}}(\omega t) \\&+ \left[1 + \mathcal{O}\left(1\over\omega^2\right)\right]J'_{\ell + {1\over2}}(\omega t)
  \\&\sim \cos\left[\omega t - {\ell \pi \over 2} + \arctan\left(\Theta + \mathcal{O}\left(1\over \omega^3\right)\over 1 - {\ell(\ell+1) \over 2\omega t}\Theta + \mathcal{O}\left(1\over \omega^2\right)\right)\right];
  \\\Theta &= {1\over\omega}\left[A_0(t) + \ell(\ell+1)A_\ell(t)\right].\label{eq:arctan}
\end{aligned}
\]
The analysis of \citet{tassoul_asymptotic_1994} also uses the same small-angle approximation. As such, we see that the derivation of \cref{eq:roxburgh58}, and thus the use of \cref{eq:tassoul}, requires that \textcircled{$\star$} there exists some location --- or ``matching point'' between the inner and outer solutions --- situated far from both the centre and surface of the star, at which \(\Theta \ll 1\). If so, then the small-angle approximation required to obtain their expressions from the full expression \cref{eq:arctan} is valid at least at this matching point, and the eigenvalue equation arising from matching them may then accurately describe the small separation.

\begin{figure}
\centering
\includegraphics[width=0.9\linewidth,height=\textheight,keepaspectratio]{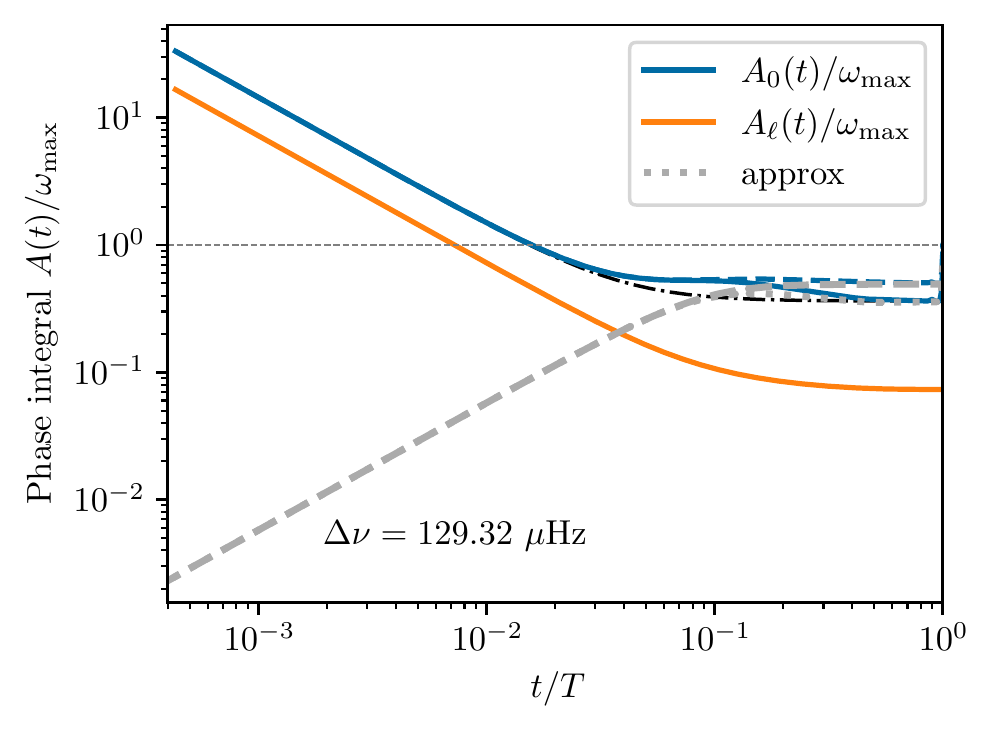}
\includegraphics[width=0.9\linewidth,height=\textheight,keepaspectratio]{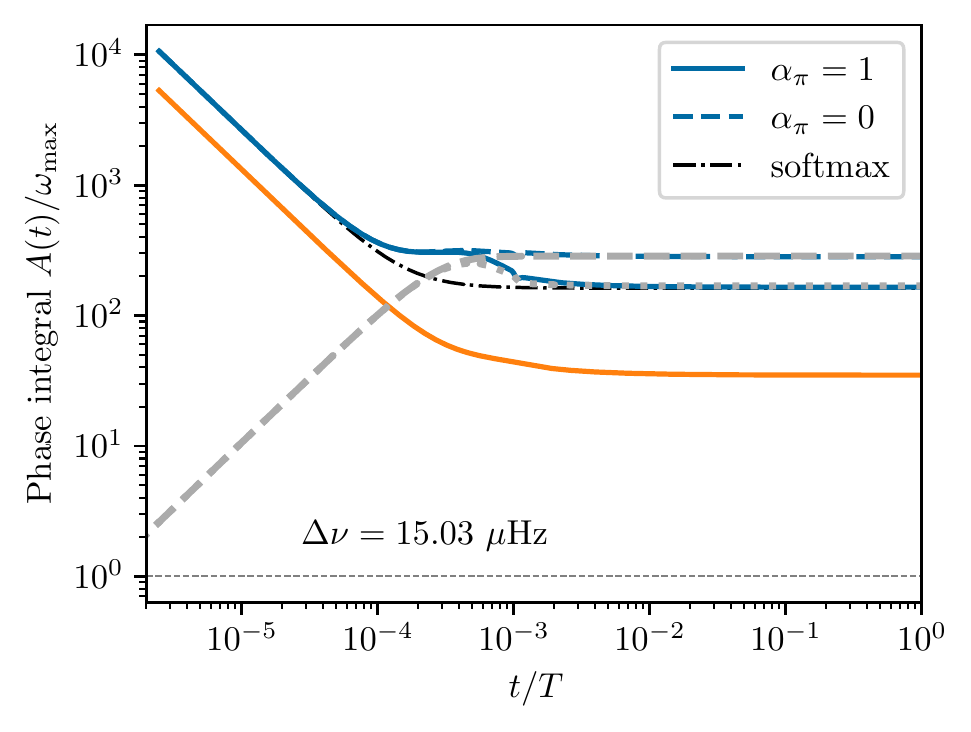}
\caption{Phase integrals \(A_0(t)\) and \(A_\ell(t)\) as defined by \rv, \cref{eq:integrals}, shown in units of \(\omega_\text{max} = 2\pi\numax\) and as functions of the acoustic radial coordinate \(t\), for two illustrative \mesa~stellar models along the same evolutionary track as shown in \cref{fig:catastrophe}. A main-sequence stellar model is shown in the upper panel, and a red-giant model is shown in the lower panel. Solid curves show values computed for the full pulsation equations (\(\alpha_\pi = 1\)), while dashed curves show values computed for the isolated \(\pi\)-modes (setting \(\alpha_\pi = 0\)). Unity is marked out with a horizontal dashed line in both panels. The small-angle approximation relied upon to derive \cref{eq:tassoul} requires that, far from the centre, both quantities \(\ll 1\); this can be seen to be satisfied for the main-sequence star, but not for the red giant. The indefinite integral used to approximate \(A_0(T)\) in \cref{eq:Iapprox} is shown with the gray curves: dotted for \(\alpha_\pi = 1\) and dashed for \(\alpha_\pi = 1\). The dash-dotted black curve shows the behaviour of our softmax approximation to \(A_0(t)/\omega\), \cref{eq:softmax}; it tends to \(1/t\) as \(t \to 0\), and to the limiting value \(A_0(T)/\omega\) as \(t\to T\).\label{fig:Aint}}
\end{figure}

To assess the validity of this assumption, we show with the solid curves in \cref{fig:Aint} the values of \(A_0(t)/\omega\) and \(A_\ell(t)/\omega\) computed from two representative stellar models along the evolutionary track shown in \cref{fig:catastrophe}: one on the main sequence (upper panel), and one on the red giant branch (lower panel). Here \(\omega\) is taken to be \(\omega_\text{max} = 2\pi \numax\). These quantities become much smaller than unity far from the center of the main-sequence model: a matching point exists for it that satisfies the condition \textcircled{$\star$} above. However, this is not the case for the red giant model: both \(A_0(t)/\omega\) and \(A_\ell(t)/\omega\) can be seen to be orders of magnitude larger than unity throughout all of the stellar structure. We submit that it is, specifically, the failure of this small-angle approximation everywhere in the stellar interior that renders \cref{eq:tassoul} unsuitable for use off the main sequence.

If this small-angle approximation should fail, the term in the denominator of \cref{eq:arctan} proportional to \(\Theta\) may also not be assumed to vanish, despite strictly speaking being of order \(1/\omega^2\). It may, however, be separated from the numerator by applying an angle addition formula, which ultimately gives
\[\begin{aligned}
   \delta_\ell(\omega, t) \sim& \arctan \left[{A_0(t)\over\omega} + \ell(\ell+1)\left({A_\ell(t)\over \omega} - {1 \over 2 \omega t}\right)\right] \\&+ \arctan\left[\ell(\ell + 1) \over 2 \omega t\right] - {\pi \over 2}\label{eq:newdelta}
\end{aligned}\]
as a leading-order asymptotic expression for the inner phase function of \cref{eq:psi}, which remains valid even where the small-angle approximation does not. Taking finite differences in the usual fashion, we then obtain that
\[
  r_{\ell, \ell + 2} \sim {2(2\ell + 3) \over \pi}\left[{a_\ell(T)/\omega\over 1 + \left[A_0(T) + \ell(\ell+1)a_\ell(T)\right]^2/\omega^2} + {1 \over 2 \omega T}\right],\label{eq:newr}
\]
where
\[
\begin{aligned}
  A_0(t) &= -{1\over2}\int_0^t \left(V_0(t') - {2 \over t'^2}\right)\mathrm d t' + {1\over t}, \text{ and}\\
  A_\ell(t) &= -{1\over2}\int_0^t \left({c^2 \over r^2} - {1\over t^2}\right)\mathrm d t' + {1\over 2t}\equiv a_\ell(t) + {1\over2t}.\label{eq:integrals}
\end{aligned}
\]
\cref{eq:tassoul} can be seen to be recovered by setting the denominator of the first term in \cref{eq:newr} to unity in the limit \(A_0(T) + \ell(\ell+1)a_\ell(T) \ll 1\) (and integrating \(A_\ell\) by parts). Importantly, for red giants in the opposite limit of large \(A_0(T) \gg a_\ell(T) \gg 1\), we see that the small separation goes simply as \(r_{02} \sim 1/\omega T \sim \Dnu /\numax\), or equivalently \(\delta\nu_{02} \sim \Dnu^2/\numax\). Here \(V_0\) is an acoustic potential (modified from eq. 59 of \rv), which we will write as
\[
  V_0(t) = \alpha_\pi N^2(t) - 4\pi G\rho + {\mathrm d w\over \mathrm d t} + w^2; \ \ \ w = {\mathrm d \over \mathrm d t} \log \sqrt{c_s \over r^2 h},\label{eq:potential}
\]
with \(G\) being the gravitational constant. The coefficient \(\alpha_\pi\) is set to 1 when analysing the usual pulsation equations, and set to 0 for the modified equations that yield the isolated ``\(\pi\)-modes'' of \citet{ong_semianalytic_2020}.

Let us now compare the estimates for the small separation as computed from our modified expressions, \cref{eq:newdelta,eq:newr}, against reference ``true'' values computed directly from the mode frequencies of stellar models along the evolutionary track shown in \cref{fig:catastrophe}. We first compute both \(r_{02}\) and \(\delta\nu_{02}\) from the mode frequencies of each model, evaluated using the \gyre~pulsation code, as functions of radial order (and thus frequency) using \cref{eq:smallsep,eq:ratio}. For the quadrupole modes, we avoid the effects of mode mixing by computing their frequencies using the \(\pi\)-mode isolation scheme of \citet{ong_semianalytic_2020}, including the first-order correction required to recover the associated pure p-mode frequencies from diagonal elements of the perturbation matrix. To provide a single numerical value for each stellar model, we then average both \(\delta\nu_{02}(\nu)\) and \(r_{02}(\nu)\), separately treated as functions of frequency, over modes near \(\numax\); we do this using a weighted sum over a Gaussian envelope centered at \numax, with a full width at half maximum \(\Gamma \sim 0.66(\numax/\mu\mathrm{Hz})^{0.88}\ \mu\mathrm{Hz}\) \citep{mosser_characterisation_2012}. We show these in \cref{fig:catastrophe}, as a function of the large separation \(\Dnu\), with the blue curves.

We compare these numerical ``ground-truth'' values with estimates of \cref{eq:newdelta,eq:newr}, shown with the light gray solid curves, when used to analyse the full pulsation equations (i.e.~with \(\alpha_\pi\) set to 1). Moreover, we show the limiting value \(3/\pi\omega T\) with the dashed curve. These can be seen to be in far better agreement with the ground truth everywhere on the evolutionary track, compared to \cref{eq:tassoul}, but nonetheless to exhibit significant remaining morphological differences.

A priori, one might attribute these remaining differences in morphology either to the truncation of the asymptotic expansion to only leading order, or to the failure of some other, similar, approximation made in its derivation. However, the only other place where the assumption of a convergent series expansion has been made by \rv~is that \(\omega \gg N\) in the stellar interior --- see property (2) in \autoref{asymptotic-analysis} --- such that the matrix \(\mathbf{C}\) enters into higher orders of asymptotic analysis by expanding \(1/\left(1 - N^2/\omega^2\right)\) in successive powers of \(N^2/\omega^2\) (see their eqs. 10 and 16). Although this approximation is locally invalid for the full pulsation equations near the centre of red giants, it is exact when we set \(\alpha_\pi = 0\) to obtain \(\pi\)-modes per the prescription of \citet{ong_semianalytic_2020}, and so may be used for analysing \(\pi\)-mode small separations without difficulty. In turn, the frequencies of these \(\pi\)-modes are known to approach those of p-modes with increasing evolution up the red giant branch.

We therefore show using the dark gray solid curves in \cref{fig:catastrophe} the values of these integral estimators,
when adapted to the \(\pi\)-mode system of equations by setting \(\alpha_\pi = 0\). The main morphological differences between the ground truth and the asymptotic estimators can be seen to persist in panels b and c, and thus cannot be attributed to this other instance of a power-series expansion failing to converge. By elimination, we must attribute them to us having truncated other higher-order terms from the asymptotic expansion. However, because these higher-order terms in the asymptotic treatment of the full p-mode system of equations are associated with increasing powers of \(N^2/\omega^2\), we also conclude that the shapes of these C-D diagrams, and their departure from smooth evolution over time, may be attributed to structural features in the Brunt-Väisälä frequency.

\cref{eq:newdelta,eq:newr} suggest that \(A_0(T)\), in addition to only \(A_\ell(T)\) as appearing in \cref{eq:tassoul}, will be required to fully describe the small separation at second order, and a succinct approximation to it will be helpful. Per \cref{eq:integrals,eq:potential}, \(A_0(t)\) is the integral of the acoustic potential \(V_0\) and other terms that are dominated by it as \(t \to T\). The acoustic potential itself contains multiple terms, and the terms proportional to \(w \sim 1/t\) are also dominated by \(\alpha_\pi N^2\) and/or \(4 \pi G \rho\) as \(t \to T\). Thus, we may approximate \(A(T)\) as
\[A_0(T) \sim - \lim_{t \to T}{1\over2}\int_0^t\left(\alpha_\pi N^2 - 4 \pi G \rho\right)\ \mathrm d t'. \label{eq:Iapprox}\]
We make this limit explicit to emphasise that, being a description of the inner phase function, numerical evaluation of this integral should in principle (1) tend to this limiting value well away from the surface of the stellar model, and (2) avoid the regular singular point at the outer boundary of the pulsation problem. We plot the values of the indefinite integral in this expression with the gray curves in \cref{fig:Aint}, shown dotted for \(\alpha_\pi = 1\) and dashed for \(\alpha_\pi = 0\). As required, its limiting values as \(t \to T\) are indeed a good approximation to those of \(A(t)\) in both cases.

Finally, in our subsequent discussion it will also be helpful to have an analytic approximation to the inner phase function \(\delta_0(\omega, t)\) for radial modes in particular. From \cref{eq:integrals,fig:Aint}, we see that \(A_0 \sim 1/t\) as \(t \to 0\), and \(\to A_0(T)\) as \(t \to T\). Thus, we approximate \(\delta_0(\omega, t)\), particularly in the limit of \(t \to 0\), as
\[
\delta_0(\omega, t) \sim \arctan\left[A_0(t)\over\omega\right]-{\pi\over2} \sim \arctan\left[\text{softmax}\left({1\over\omega t}, \tan\beta\right)\right]-{\pi\over2},\label{eq:softmax}
\]
where we have defined \(\tan(\beta) \equiv A_0(T)/\omega\) so that \(\delta_0(t)\to\beta - {\pi/2}\) as \(t\to T\). The softmax function (often used in the statistical and machine-learning literature) is some smooth approximation to the maximum function, which returns the greater of its two arguments. A natural choice for implementing such smoothing is the \(L^2\) norm, such that \(\text{softmax}(a, b)\sim\sqrt{a^2+b^2}\). We show the behaviour of this approximation for \(A_0(t)\) using the black dash-dotted lines in \cref{fig:Aint}. Its shape describes that of \(A_0(t)\) reasonably with \(\alpha_\pi\) set to 1, and quite well with \(\alpha_\pi\) set to 0.

\section{Intepreting C-D Diagrams}\label{intepreting-c-d-diagrams}

\begingroup
\renewenvironment{figure}{\begin{figure*}}{\end{figure*}}

\begin{figure}
\centering
\annotate{\includegraphics[width=0.9\linewidth,height=\textheight,keepaspectratio]{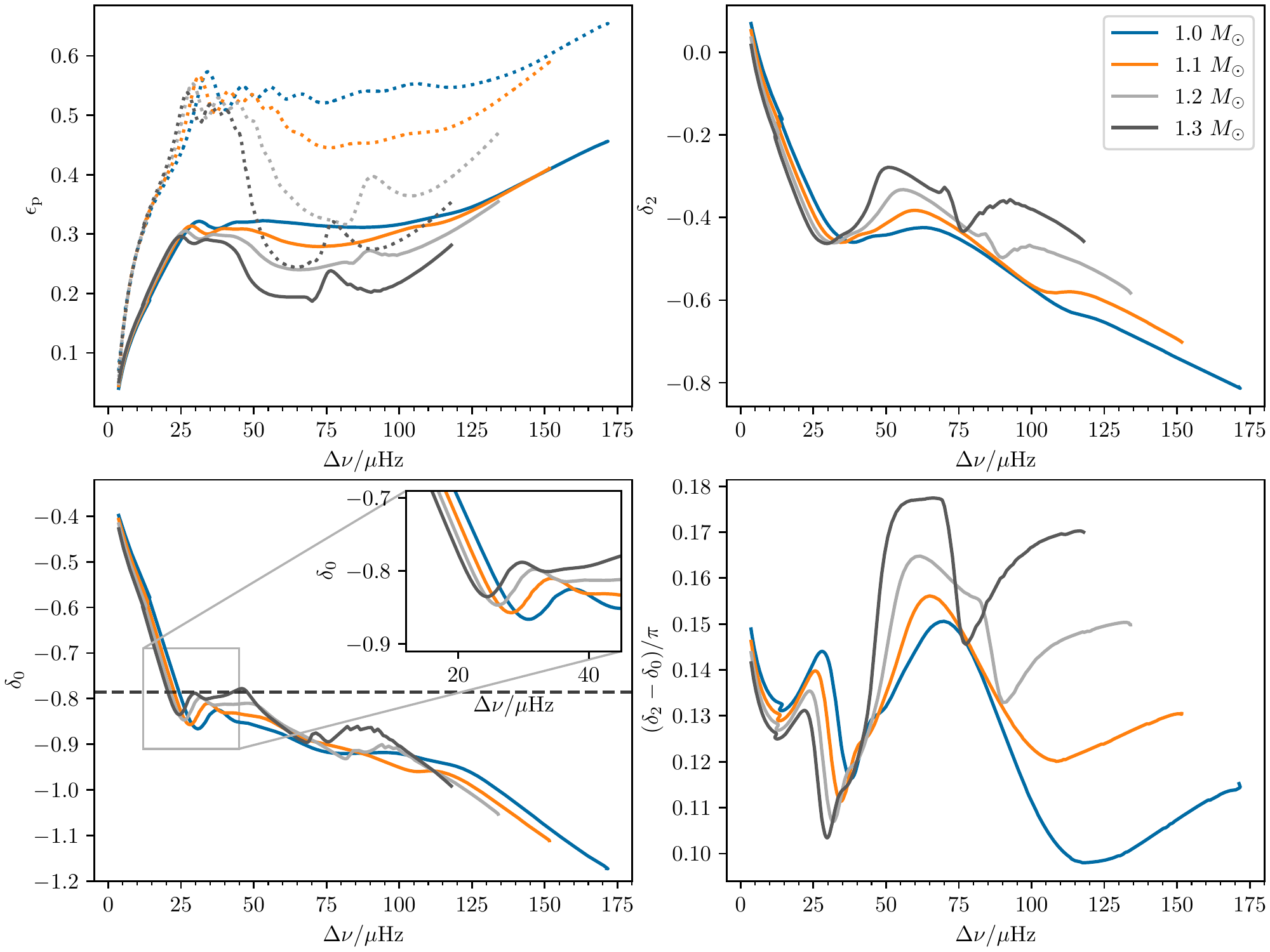}}{\node at (.1, .94){\textbf{(a)}};\node at (.1, .12){\textbf{(c)}};\node at (.8, .94){\textbf{(b)}};\node at (.95, .12){\textbf{(d)}}; \draw[black, thick](.3, .64) -- (.35, .64); \draw[black, thick, dotted](.3, .605) -- (.35, .605); \node[right] at (.35, .64){$\nu_{n0}/\left<\Delta\nu\right> - n$}; \node[right] at (.35, .605){$2\nu_{n0} T - n$};}
\caption{The evolution of several seismic diagnostic quantities off the main sequence and up the red giant branch, as computed from numerical frequency calculations on evolutionary tracks of \mesa~stellar models. Tracks of stellar models with different masses are indicated with solid curves of different colours. \textbf{(a)}: The radial-mode phase offset \(\epsilon_\mathrm{p}\). Solid curves show theoretical values, where \(\epsilon_\mathrm{p}\) is derived as an offset from integer multiples of the asymptotic frequency spacing \(1/2T\), while dashed curves show mock observational values, computed with respect to a mean separation \(\left<\Dnu\right>\). The mock observational values are offset by \(-0.8\), and both the theoretical and mock observational values are plotted using the mean separation \(\left<\Dnu\right>\) as the abscissa. \textbf{(b)}: The quadrupole-mode inner phase function evaluated at \(\numax\), \(\delta_2\left(\omega_\text{max}\right)\). These curves can be seen to evolve smoothly in the neghbourhood of the bottleneck in \(\epsilon_\mathrm{p}\) at \(\Dnu\sim 25 \mu\mathrm{Hz}\). \textbf{(c)}: The radial-mode inner phase function evaluated at \(\numax\), \(\delta_0\left(\omega_\text{max}\right)\). Unlike \(\delta_2\), small undulatory features can be seen in the neighbourhood of the bottleneck, which we show more clearly in the inset figure. Roughly speaking, the value of \(\delta_0\) here is about \(-\pi/4\), which we mark out with the horizontal line. \textbf{(d)}: The difference between the two, scaled to yield an estimate of the small separation ratio \(r_{02}\) by \cref{eq:phasediff}. The shape of the Knee feature reported in \citet{reyes_in_review} arises from variations in \(\delta_0\) alone, of the kind shown inset in panel (c). \label{fig:evol}}
\end{figure}
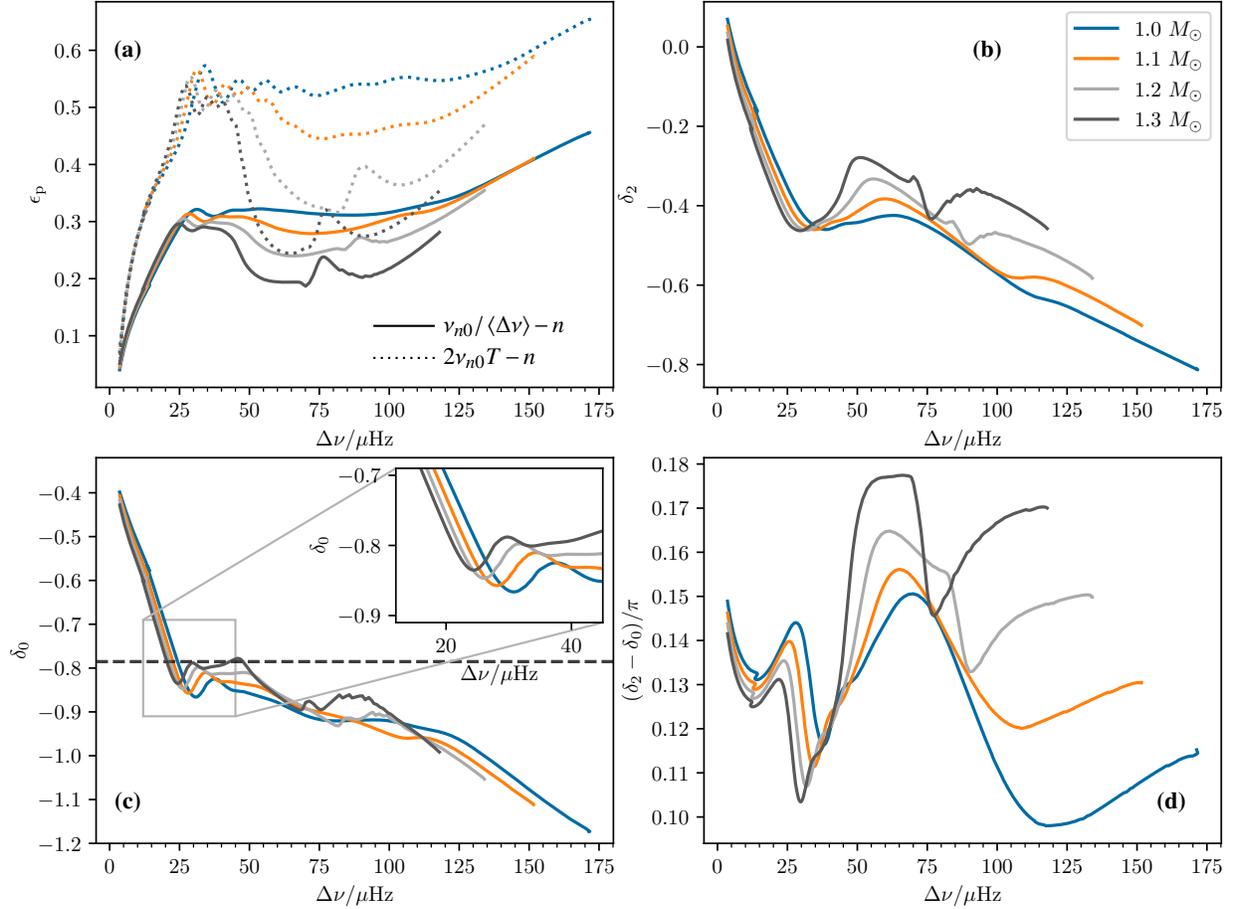

\endgroup

All of our preceding analysis has been restricted to the asymptotic behaviour of the eigenfunctions, with comparisons to numerical mode frequencies serving only to assess the properties and limitations of this asymptotic analysis. We now consider the problem of interpreting both these numerical calculations, and observational measurements, using this asymptotic analysis as a guide.

Our discussion here is motivated by \citet{white_diagrams_2011}, who described undulatory features in the \(\epsilon_\mathrm{p}-\Dnu\) diagrams calculated using the radial-mode frequencies from evolutionary tracks of stellar models. \citet{ong_structural_2019} used similar calculations to generate isochrones on the \(\epsilon_\mathrm{p}-\Dnu\) plane, in which these features remain visually apparent. One feature of these calculations was a bottleneck at \(\Dnu\sim20-30\ \mu\mathrm{Hz}\), after which both the evolutionary tracks and the isochrones converge to a single sequence on this plane (which we show in \cref{fig:evol}a). Intuitively, this makes sense because very evolved red giants (at low values of \(\Dnu\)) are almost fully convective, and so both their \(\Dnu\) and \(\epsilon_\mathrm{p}\) are primarily determined by homologous scaling against convectively stratified polytropes \citep{mosser_universal_2011}. For these stars, additional phase perturbations from the radiative core \citep[e.g.][]{ong_semianalytic_2020}, near-surface effects \citep[e.g.][]{li_surface_2023}, and acoustic glitches \citep[e.g.][]{dreau_glitch_2020, saunders_glitch_2023} are all relatively small. By contrast, these same structural features more significantly affect the mode frequencies of less-evolved stars. This bottleneck signifies a transition between these two regimes.

However, the precise details of this transition are not well-studied, with a historical lack of attention to it owing to a paucity of observational data. For example, \(\numax\) for stars in this regime of evolution lie in the gap between long- and short-cadence samples from the nominal \emph{Kepler} mission. However, in light of recent observational findings, a proper theoretical treatment of it has now become necessary. In particular, \citet{reyes_in_review} produce the first empirical C-D diagram derived from measurements of small and large separations in a single coeval stellar population --- the pulsating sub- and red giants of the open cluster M67. Their main result is the observational discovery of a ``knee'' in this cluster C-D diagram (or, equivalently, a ``hump'' in the \(r_{02}-\Dnu\) diagram). These features are highlighted in the red boxes on the evolutionary tracks depicted in \cref{fig:catastrophe}. Strikingly, this knee also occurs at \(\Dnu\sim20\ \mu\mathrm{Hz}\), coinciding with the transition of p-modes from being determined primarily by structure, to being determined primarily by homology.

The ``knee'' feature in the C-D diagram (which we will subsequently refer to as ``the Knee'' for brevity) had not been noted in earlier observational studies, most of which combined multiple stellar populations, owing to the observational selection effects that we have discussed earlier. It also has not previously been characterised by numerical studies. \citet{ong_structural_2019} restricted their attention to diagnostics derived from radial modes, while the calculations of \citet{white_diagrams_2011}, which predates the derivation of \(\pi\)-mode isolation schemes in \citet{ball_surface_2018} and \citet{ong_semianalytic_2020}, were dominated by numerical scatter from gravitoacoustic mode coupling, which made the Knee difficult to notice. With the benefit of these new numerical techniques, however, \citet{reyes_in_review} were able to reproduce it in their evolutionary modelling. Finding \cref{eq:tassoul} not to be usable in interpreting this feature, for the reasons described in the previous section, \citet{reyes_in_review} turned to \cref{eq:phasediff} instead, to interpret their evolutionary calculations.

We show the results of similar calculations in detail in \cref{fig:evol}, computed with respect to illustrative evolutionary tracks. The ones shown here were generated with \mesa~r22.05 \citep{mesa_paper_1, mesa_paper_2, mesa_paper_3, mesa_paper_4, mesa_paper_5, mesa_paper_6}, and were calculated using the default equation of state and opacity tables, adapted to the chemical mixture of \citet{GS98}, at solar metallicity, using a solar-calibrated helium abundance and mixing length, an Eddington-gray atmospheric boundary condition, with element diffusion and gravitational settling using the formulation of \citet{thoul_diffusion_1994}, and a mass-dependent diffusion scaling prefactor per the prescription of \citet{viani_metallicity_1994}, but without radiative levitation, rotational mixing, or turbulent diffusion or pressure support. These specific evolutionary tracks shown were also generated using convective envelope overshooting, with \(f_\text{ov} = 0.08\), and with a finer radial coordinate mesh than default\footnote{In order to better resolve the evolution of the stellar model's convective boundaries, we used a mesh 4 times denser than default, and set the \texttt{convective\_bdy\_weight} parameter to 5. These options necessitated increasing the maximum number of allowed model zones to \(20,000\).}. In this figure, we vary only the stellar mass, keeping all other quantities fixed. However, we note that our objective is not to study the evolutionary, compositional, or other parametric dependences of the Knee feature per se, but rather only to examine how it relates to the internal structure of the stellar models under consideration.

The inner phase shifts \(\delta_0\) and \(\delta_2\) for each of the models in our tracks were then calculated from each model's oscillation frequencies and eigenfunctions, computed using \gyre. The radial p-mode and quadrupole \(\pi\)-mode frequencies and eigenfunctions were calculated within \(\pm 7.5 \Delta \nu\) of \(\nu_{\text{max}}\); the latter according to the \(\pi\)-mode isolation construction of \citet{ong_semianalytic_2020}. Per \cref{eq:psi}, the inner phase shift of a particular mode, \(\delta_{\ell}\), as a function of the acoustic radius, is calculated by evaluating
\[\begin{aligned}
\delta_{\ell}(\omega, t) &\sim \arctan\left(\psi \left/{\mathrm d \psi \over \mathrm d (\omega t)}\right.\right) - \omega t \\
&= \arctan \left( \frac{\psi \cos(\omega t) - \frac{\mathrm d \psi}{\mathrm d (\omega t)} \sin(\omega t) }{\frac{\mathrm d \psi}{\mathrm d (\omega t)} \cos(\omega t) + \psi \sin(\omega t)} \right).
\end{aligned}\]
We choose to evaluate \(\delta_0\) and \(\delta_2\) at \(t = T/2\). In order to get a single value of \(\delta_0\) and \(\delta_2\) for each model, we compute the weighted average over all modes with respect to a Gaussian envelope, whose full-width \(\Gamma\) is specified by the same formula of \citet{mosser_characterisation_2012} as used in the preceding section.

Decomposing the separation ratio \(r_{02}\) into contributions from the quadrupole (\cref{fig:evol}b) and radial (\cref{fig:evol}c) modes, as in our figure, \citet{reyes_in_review} find that the Knee (easily visible in \cref{fig:evol}d) is associated with localised structure in the radial-mode phase function alone. Correspondingly, they concluded that, whatever the origin of this structural feature, it must lie so deep in the stellar interior that only radial p-modes are able to probe it --- in other words, it is closer to the centre than the Jeffreys-Wentzel-Kramers-Brillouin (JWKB) inner turning point of the quadrupole modes. They put forward the hypothesis that this structural feature is the acoustic glitch at the convective envelope boundary, and moreover demonstrate a qualitative correspondence between the morphology of the Knee, and the numerical values of the frequency-response kernel for the density \(K_{\rho, c_s}\) at the convective boundary, for stellar models along an isochrone passing through the Knee.

\subsection{Relating Acoustic Glitches to the C-D Diagram}\label{relating-acoustic-glitches-to-the-c-d-diagram}

Our preceding theoretical construction now positions us to interpret this phenomenon analytically. In particular, our revised asymptotic analysis permits us to overcome certain fundamental limitations in the qualitative analysis presented in \citet{reyes_in_review}. For one, the kernels \(K_{\rho, c_s}\) and \(K_{c_s,\rho}\) describe how numerical mode frequencies would respond to localised perturbations \(\delta\rho\) and \(\delta c_s\), respectively, to the existing stellar structure. However, in the context of acoustic glitches, the frequency differences of interest are those induced relative to some notional glitch-free stellar structure, where the mode frequencies are primarily determined by large-scale features of the acoustic mode cavity. It is precisely this behaviour that is captured by low-order asymptotic analysis of the kind that we have examined earlier. In this sense, low-order asymptotic approximations to frequency-response kernels, which we will build, are therefore more appropriate for the study of acoustic glitches specifically, compared to the purely numerical ones ordinarily used in structure inversions \citep{gough_inversion_1991, kosovichev_inversion_1999}. Moreover, the pair of kernels in the standard variables \((\rho, c_s)\) showed additional features that \citet{reyes_in_review} were unable to associate with those of the observed C-D diagram.

\begin{figure}
\centering
\pandocbounded{\includegraphics[keepaspectratio]{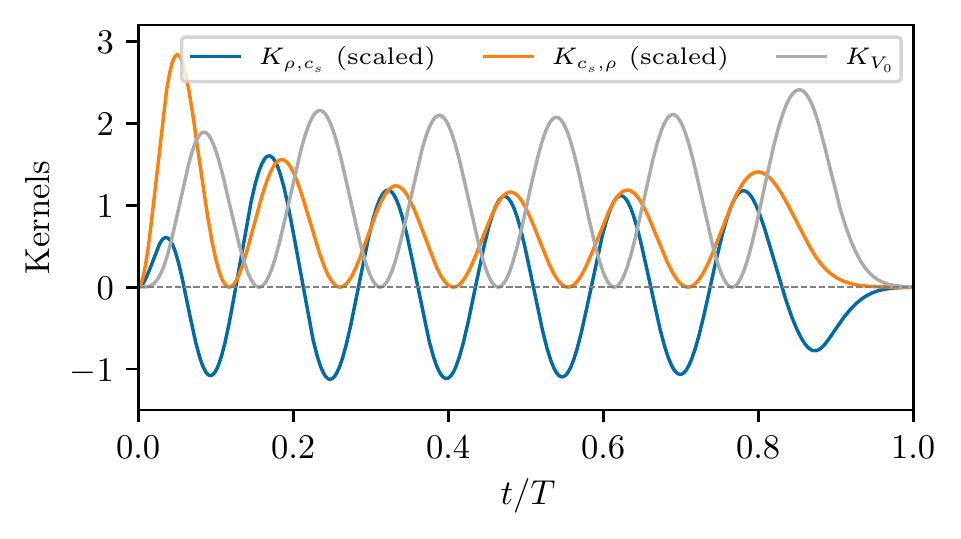}}
\caption{Comparison of frequency-response kernel to the acoustic potential \(V_0\), \(K_{V_0}(t)\), against the structure inversion kernels \(K_{c_s, \rho}\) and \(K_{\rho, c_s}\), all computed for the radial p-mode at \(n_p = 6\) of a main-sequence stellar model. The latter two are scaled by \(c_s\) to accommodate integration against the acoustic radius \(t\) rather than the physical radius \(r\) as is customary, thereby showing near-centre features more clearly. The \(V_0\) kernel can be seen to be morphologically distinct from the usual kernel pair: it is phase-lagged in quadrature from the \(c_s\) kernel, and has nonzero overall integral unlike the \(\rho\) kernel. \label{fig:kernelcomparison}}
\end{figure}

We will now derive a different frequency-response kernel whose features match more closely with the observational diagram. Given our above discussion, we restrict our attention to understanding how near-core structural features may affect the frequencies of radial modes in particular. For radial modes only, the pulsation equations reduce to a second-order problem in the canonical form \citep{gough_linear_1993}
\[
  {\mathrm d^2 \over \mathrm d t^2}\psi + \left(\omega^2 - V_0(t)\right)\psi = 0,\label{eq:canonical}
\]
expressed in terms of \(\psi = \xi_r\sqrt{r^2 \rho c_s}\) and the acoustic radial coordinate \(t\). Per \cref{eq:potential}, the acoustic potential \(V_0(t)\) can be seen to depend on both \(N\), and potentially its derivatives. Following standard results in perturbation theory \citep{houdek_asteroseismic_2007}, any localised departures from an otherwise smooth acoustic potential, or perturbations \(\delta V(t)\) to it, result in corresponding perturbations to the mode frequencies of the form
\[
  \delta\omega^2_n \sim \int \left|\psi_n\right|^2 \delta V(t)\ \mathrm d t \equiv \int K_n(t) \delta V(t)\ \mathrm d {(t/T)},
\]
where the eigenfunctions are by convention each of unit norm, so that the kernels \(K_n\) are of unit integral. As can be seen in \cref{fig:kernelcomparison}, these kernels in \(V_0\) are morphologically distinct from the structural kernels \(K_{c_s, \rho}\) and \(K_{\rho, c_s}\) often used to characterise the response of mode frequencies to structural perturbations in seismic inversions \citep[e.g.][]{gough_inversion_1991, kosovichev_inversion_1999}, or to describe the behaviour of acoustic glitches \citep[e.g.][]{mazumdar_glitch_2014, verma_glitch_2014, verma_seismic_2017, reyes_in_review}. Because the mode frequencies are uniquely determined by \(V_0\) (and boundary conditions), there is no cross-term kernel.

At \(\ell = 0\), the Riccati-Bessel function \(s_{\ell = 0}(x)\) is just \(\sin x\). Neglecting the variation in the amplitude function \(A(t)\), \cref{eq:psi} then gives
\[
  K_n(t) \sim 2 \sin^2 \left(\omega_n t + \delta_0(\omega_n, t)\right),
\]
with unit normalisation recovered in the JWKB limit of high \(n\). As \(\delta(\omega^2_n) \sim 2\omega_n \delta\omega_n = 8\pi^2\nu_n\delta\nu_n\), perturbations to the phase offset \(\delta\epsilon_n = \delta\nu_n / \Delta\nu\) may also be related to such integral kernel expressions.

\begin{figure}
\centering
\pandocbounded{\includegraphics[keepaspectratio]{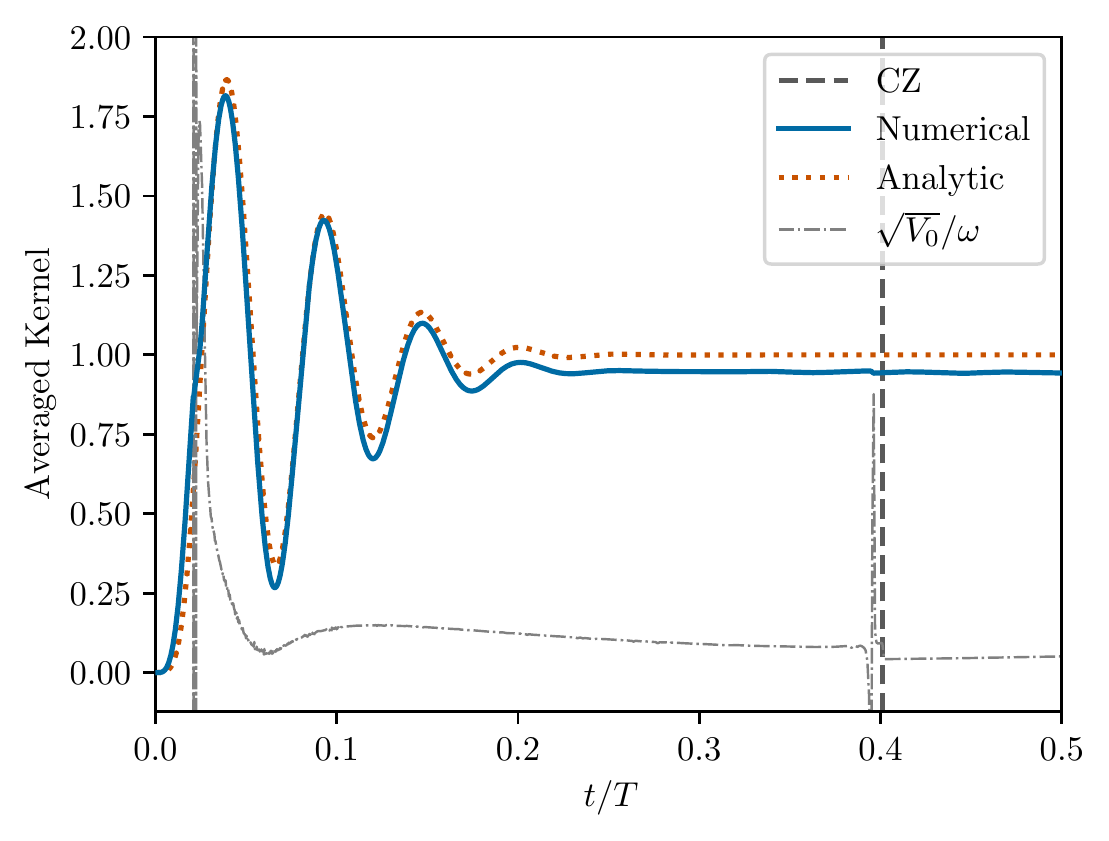}}
\caption{Overview of structural features near the center of an evolutionary series of stellar models. The averaged kernel \(K(t) = \sum_n w_n K_n(t)\) is shown with the solid blue curve, for an illustrative stellar model. The acoustic potential \(V_0\) --- which fully determines the structure of radial modes through \cref{eq:canonical} --- is shown with the dash-dotted curve, while the location of the convective envelope boundary is indicated with the vertical dashed line. Our analytic approximation to the averaged kernel, \cref{eq:K}, is shown with the dotted line, as computed with \(\beta = {\pi \over 4}\). \label{fig:singlekernel}}
\end{figure}

\begin{figure}
\centering
\annotate{\pandocbounded{\includegraphics[keepaspectratio]{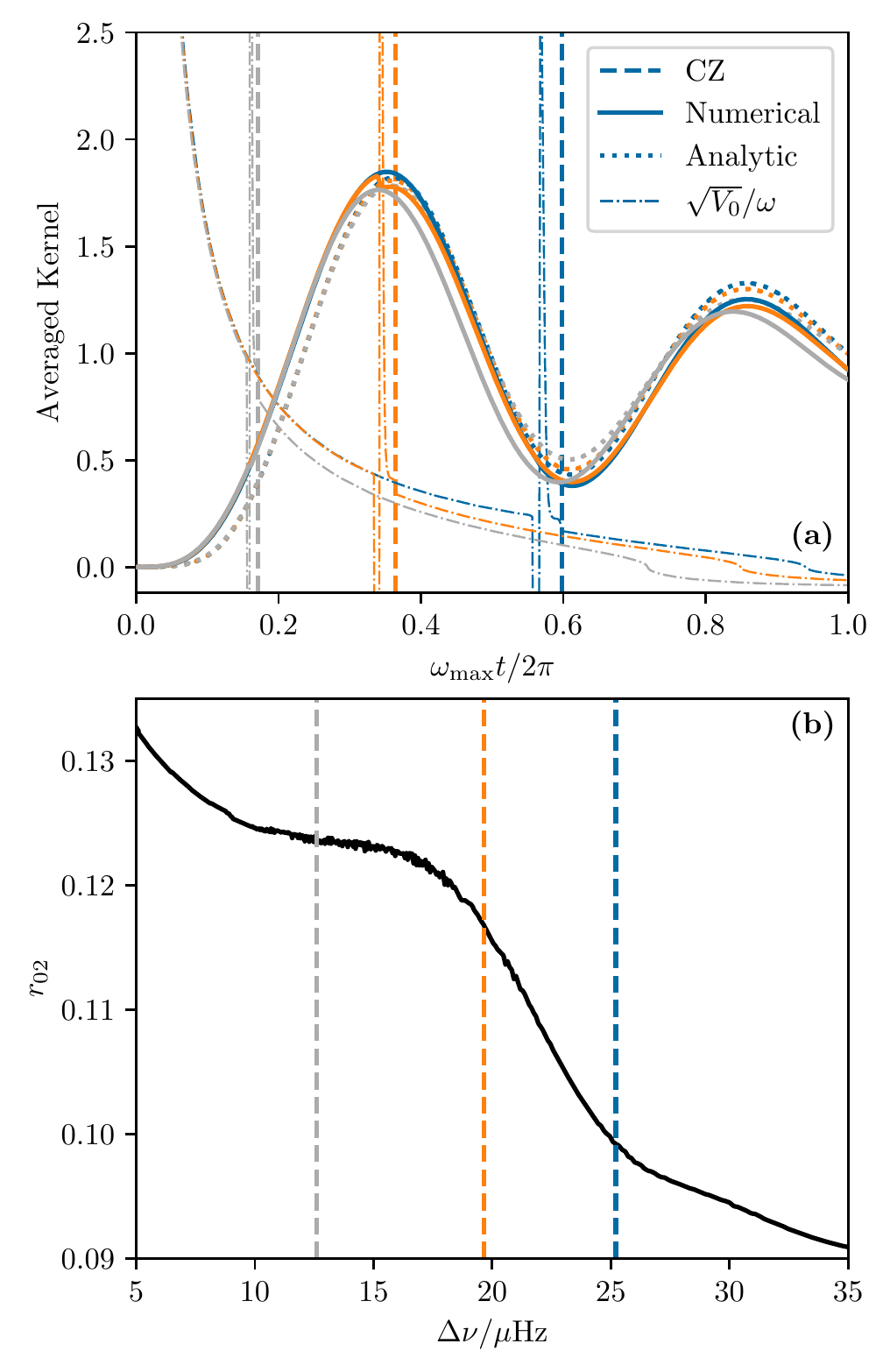}}}{\node[rotate=-23] at (.815,.15){\footnotesize$\leftarrow$ time evolution};}
\caption{Detailed examination of the correspondence between quantities shown in \cref{fig:singlekernel}, and their signature in a C-D diagram. \textbf{(a)}: Evolution of the convective envelope boundary during the ``hump'' feature in the \(r_{02}\) diagram. Three stellar models are shown with curves of different colours, with each linestyle possessing the same meaning as indicated in \cref{fig:singlekernel}. Over the course of evolution between these snapshots, the convective envelope boundary passes through the innermost maximum of the averaged kernel. \textbf{(b)}: The \(r_{02}\) diagram associated with this evolutionary track, with the stellar models depicted in panel (a) indicated using vertical dashed lines of the same colours as the corresponding curves there. The animated version of this figure shows the continuous time evolution of both panels from the model shown in blue to that shown in gray, over a duration of 6 seconds, during which the convective boundary sweeps over the innermost maximum of the averaged kernel (the static figure providing three snapshots of this process). \label{fig:kernelszoom}}
\end{figure}

While this suffices for computing frequency and phase perturbations on a mode-by-mode basis, these phase offsets and ratios are, in observational practice, often measured by way of averaging over many modes \citep[using, say, the ``collapsed-échelle-diagram'' technique, as in][]{reyes_in_review}. The weights \(w_n\) of this average are in effect given by the observable heights of the modes in the power spectrum, which we again take to be specified by the Gaussian envelope of \citet{mosser_characterisation_2012}. With some work (details of which we provide in \autoref{sec:kernel}), we may relate this averaged phase offset to an integral over an averaged kernel,
\[
\left<\delta\epsilon\right> = \left<\delta\left(\delta_0\right)\right> \sim {1 \over 8\pi^2 \nu_\text{max} \Delta\nu} \int K(t)\ \delta V(t)\ \mathrm d (t/T),
\]
where \[\begin{aligned}K(t) &=  \sum_n w_n K_n(t) \\&\sim \left[1 - \cos (2\omega_\text{max} t - 2\delta_0(\omega_\text{max}, t))\right] \cdot \exp\left[-2 \Sigma^2 t^2\right].\end{aligned} \label{eq:Kdef}\]
In words, this averaged kernel is well approximated by a harmonic function modulated by a Gaussian envelope of width \(1/2\Sigma\) centered at \(t = 0\). We illustrate this averaged kernel using the solid curve in \cref{fig:singlekernel}, and our analytic approximation to it, \cref{eq:K}, using the dotted curve, with \(\beta \sim {\pi \over 4}\) (horizontal line in \cref{fig:evol}c).

We now possess the required analytic infrastructure to examine the ``Knee'' of \citet{reyes_in_review}. Our preceding asymptotic analysis suggests that it is caused by some structural feature of the Brunt-Väisälä frequency. This justifies the hypothesis of \citet{reyes_in_review} attributing it to the inner boundary of the convective envelope. This boundary advances towards the centre of the star over the course of evolution up the red giant branch. To illustrate this, we show the location of this convective boundary using the vertical dashed line in \cref{fig:singlekernel}. We also show the acoustic potential \(V_0\) itself using the dash-dotted line. At the location of the convective boundary, we find a significant and highly localised feature in the acoustic potential, of the kind that might induce an acoustic glitch. However, given that the shape of the average kernel there in the depicted model is more or less uniform, the inner phase function \(\delta_0\) of the depicted model is sensitive only to the shape, rather than position, of the convective boundary. That is to say, when the boundary is far from the stellar centre, the radial-mode frequencies will collectively be insensitive to small changes to the location of the envelope boundary, all else being equal.

Let us now turn our attention to the time evolution of these quantities in stellar models spanning the Knee. We show all of the above quantities, with colours indicating different stellar models, in \cref{fig:kernelszoom}a. The corresponding locations of these models in the \(\Dnu-r_{02}\) diagram are indicated with the correspondingly coloured vertical dashed lines in \cref{fig:kernelszoom}b. Over the course of evolution through the Knee in the \(r_{02}-\Dnu\) diagram, the convective envelope boundary can be seen to sweep over the innermost maximum of the averaged kernel \(K\). As such, the amount by which \(r_{02}\) is enhanced over its asymptotic value is determined by the value of the kernel at the location of the boundary glitch in the acoustic potential. Correspondingly, the shape of the overall feature in the \(r_{02}\) diagram appears to be determined by the shape of the innermost maximum in the kernel as the acoustic glitch passes through it.

\begin{figure}
\centering
\pandocbounded{\includegraphics[keepaspectratio]{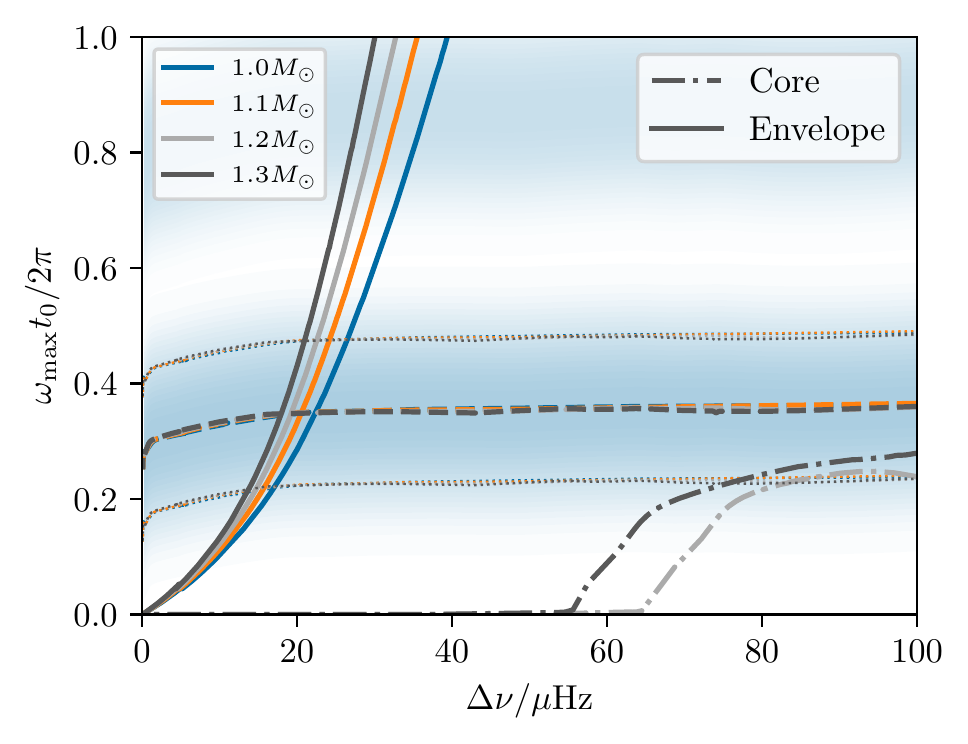}}
\caption{Time evolution of convective boundaries relative to location of the innermost maximum in the radial-mode sensitivity kernel, whose sensitivity for the evolutionary track at \(1.2 M_\odot\) is schematically indicated by the shading of the background. Solid curves show the locations of the convective envelope boundaries, while dash-dotted ones show those of the convective core, if any. The dashed curves show the location of the innermost maximum of the radial-mode averaged kernels as computed from numerical eigenfunctions, and the dotted curves around them indicate the notional half-width of \(\pm\pi/4\) radians, in the acoustic phase coordinate \(\omega_\text{max} t\). Different colours indicate different stellar masses, with the same colour coding as in \cref{fig:evol}. \label{fig:kernelevol}}
\end{figure}

Critically, because the features of the averaged kernel are modulated by a Gaussian envelope located at the centre of the star, this innermost maximum is where the radial modes are collectively \emph{maximally} sensitive to any acoustic glitch. Its location can be found by solving for the acoustic radius \(t_0\) such that
\[\omega_\mathrm{max} t_0 + \delta_0(\omega_\text{max},t_0) = {\pi \over 2}.\]
Even using our greatly simplified approximate expression for \(\delta_0\), this is a transcendental equation in \(t_0\), with no analytic solutions. Numerically, however, the position of this innermost maximum does not appear to vary significantly in our \mesa~evolutionary calculations. We illustrate this for a series of evolutionary tracks in \cref{fig:kernelevol}. Over the range of masses that we have shown in \cref{fig:evol}, and for most of stellar evolution up to the red giant branch, the position of this maximum can be seen to remain very stable, at \(t_0 \sim 0.35/\numax\). As such, the shape of the hump feature in the \(r_{02}\) diagram may be robustly interpreted as diagnosing the location of the convective boundary over evolution, relative to this particular acoustic distance from the centre of the star. As such, whereas \citet{reyes_in_review} have shown the Knee numerically to be a sensitive diagnostic of convective overshooting, our asymptotic analysis ties it directly to a feature of stellar structure.

\section{Discussion and Conclusion}\label{discussion-and-conclusion}

The commonly-used formula for the small separation given by \citet{tassoul_asymptotic_1994}, \cref{eq:tassoul}, does not apply to red giants. By revisiting the asymptotic analysis used in its original construction, we have derived a single expression, \cref{eq:newr}, which both reduces to that of \citet{tassoul_asymptotic_1994} for main-sequence stars, and yet also reproduces the qualitative behaviour of small separations computed from numerical mode frequencies, when applied to red-giant stellar models. The transition between the two regimes is governed by the parameter \(A_0(T)\) of \citet{roxburgh_asymptotic_1994}. We derive an approximation for it, \cref{eq:Iapprox}, as an acoustic integral of a combination of the Brunt-Väisälä frequency and the local density. On the main sequence, \(A_0(T) \ll 2\pi\numax\), while \(A_0(T) \gg 2\pi\numax\) in red giants.

For sufficiently evolved red giants, the frequencies of radial modes are known to be well-approximated by homologous scaling against polytropes with condensed cores \citep[e.g.][]{gabriel_properties_1979}. Our analysis indicates that all of the pure p-modes of such red giants, rather than just the radial modes, are likewise well-approximated (ignoring coupling with the inner g-mode cavity). This is consistent with the known observational features of p-modes in very evolved red giants, where the small separation ratio does not vary significantly between stars with similar \(\Dnu\) \citep[e.g.][]{bedding_solarlike_2010, huber_asteroseismology_2010, mosser_universal_2011}, even despite differences in other stellar properties. In particular we find that, in these very advanced stages of red giant evolution, the small separation ratios only scale with \(\Dnu/\numax\), rather than yielding any diagnostic information about their internal structure as previously assumed.

\begin{figure}
\centering
\includegraphics[width=1\linewidth,height=\textheight,keepaspectratio]{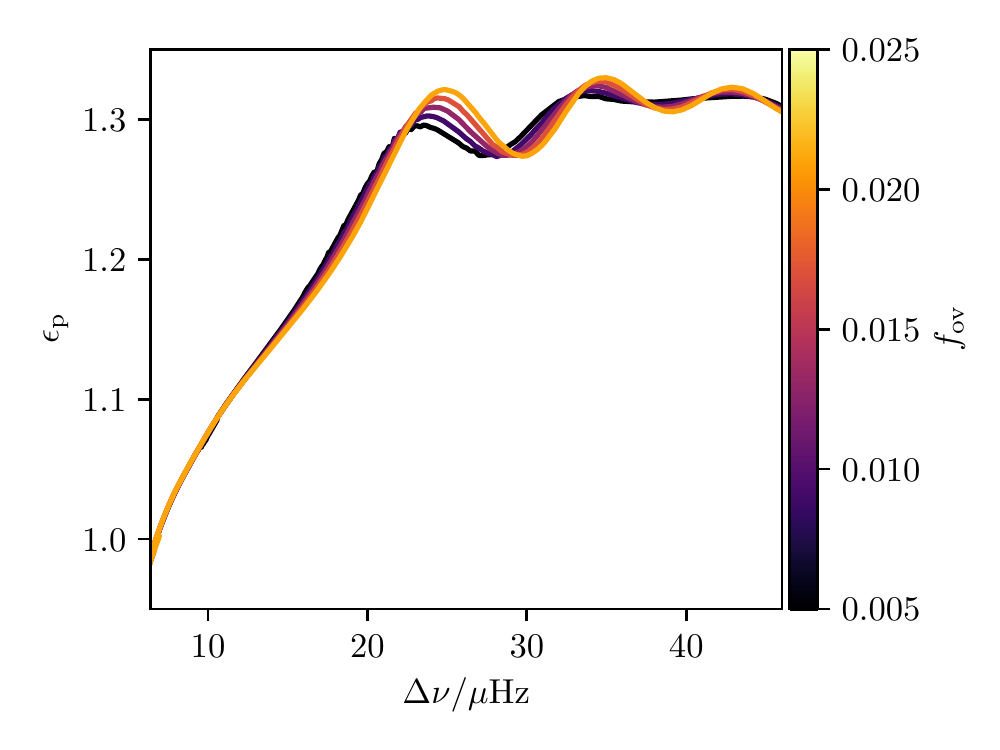}
\includegraphics[width=1\linewidth,height=\textheight,keepaspectratio]{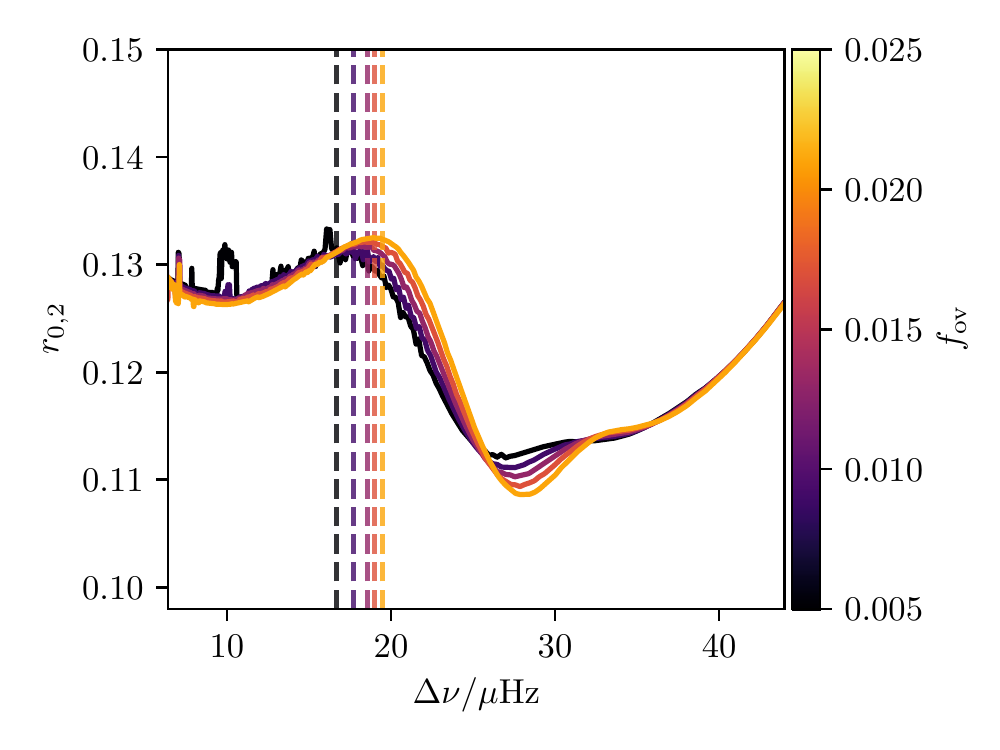}
\caption{Dependence of \(\epsilon_\mathrm{p}\) (upper panel) and \(r_{02}\) (lower panel) on convective overmixing. Both panels show these quantities as computed from evolutionary tracks at \(1.2M_\odot\), with overmixing implemented with \mesa's prescription of an exponential falloff in the diffusion coefficient, with a scale length of \(l_\mathrm{ov} = H_P f_\mathrm{ov}\). At earlier times than shown, and further up the red giant branch, these evolutionary tracks coincide; however, they separate based on the amount of overshoot applied, in the neighbourhood of the feature in \(r_{02}\). The separation in the \(\epsilon_\mathrm{p}\) diagram can be seen to precede that in the \(r_{02}\) diagram. The size of the enhancement to \(r_{02}\) does not appear to vary significantly, relative to observational uncertainties, but, as with the isochrones in \citet{reyes_in_review}, the value of \(\Dnu\) at which it attains a maximum does appear to be sensitive to the amount of overmixing in the model physics. \label{fig:overshoot}}
\end{figure}

Evolution off the main sequence to the red giant branch is also accompanied by qualitative changes in how the structural effects of the convective envelope boundary are encoded in the mode frequencies, and our asymptotic analysis now allows this transition to be understood in detail. On the main sequence, analysis of the pulsation equations in the JWKB approximation indicates that the convective envelope boundary imprints an oscillatory ``acoustic glitch'' signature into the frequencies of solar-like oscillators, which is independent of the degree \(\ell\) \citep[e.g.][]{mazumdar_glitch_2014, verma_glitch_2014, verma_seismic_2017}. By contrast, the radiative cores of red giants are extremely compact. \citet{lindsay_nearcore_2023} show that such convective boundaries, located close to the centre of the star, instead induce perturbations into the p-mode frequencies that scale as power laws with degree-dependent indices. The novel observational feature reported in \citet{reyes_in_review} --- a ``knee'' in the C-D diagram --- occurs during the transition between these two regimes. In this work, we have shown that during this transition,

\begin{enumerate}
\def\labelenumi{\arabic{enumi}.}
\tightlist
\item
  the averaged values of both the p-mode phase offset, and of the small separation, are maximally sensitive to the structure of the convective boundary (both encoding the radial-mode phase function \(\delta_0\)).
\item
  At this point, the convective envelope may be very robustly characterised as lying at an acoustic distance roughly a third of \(1/\numax\) away from the centre of the star.
\end{enumerate}

It is interesting to contrast this second result with existing techniques for characterising the envelope-boundary glitch in main-sequence stars, where it is instead the unknown acoustic position of the glitch that is the object of observational determination.

Our qualitative discussion relates the position and shape of the Knee on the C-D diagram only directly to the position of the convective boundary, and the shape of an averaged radial-mode kernel. However, both of these are ultimately, if indirectly, determined by global properties of the physical processes governing stellar structure and evolution. For example, \citet{reyes_in_review} point out that the shape of isochrones in model C-D diagrams are strongly affected by the amount of convective boundary mixing beneath the envelope (e.g., undershooting of convective velocities) used to generate these models, which determines the position of the convective boundary, all else being equal. We illustrate other quantitative effects of this on both the radial-mode phase function \(\epsilon_\mathrm{p}\), and in the surface-insensitive \(r_{02}\) diagram, in \cref{fig:overshoot}. In our stellar models and frequency calculations, convective boundary mixing modifies the undulatory features in \(\epsilon_\mathrm{p}\) that emerge preceding the envelope boundary passing through the hump in the \(r_{02}\) diagram, and modifies not just the position of the Knee, but also its overall morphology.

Thus, combined measurements of \(\Delta\nu\), \(\epsilon_p\), and \(r_{02}\) (or \(\delta\nu_{02}\) with surface correction), during this asymptotic transition, will constrain convective boundary mixing beneath convective envelopes. Should further observational efforts at this be realised, these probes would be complementary to other seismic indicators of convective boundary mixing there, such as those obtained from g-mode seismology in the lead-up to the luminosity bump \citep[e.g.][]{lindsay_bump_2022}, or from the position of the luminosity bump itself on the seismic Kiel diagram \citep[e.g.][]{khan_bump_2018}. We leave the continued observational pursuit of this feature --- either in other coeval stellar populations, or in the field --- and a fuller characterisation of its ability to probe convective boundary mixing, to future work.

\section*{Acknowledgements}\label{acknowledgements}
\addcontentsline{toc}{section}{Acknowledgements}

We thank the anonymous referee for suggestions that improved the clarity of this work, Earl Bellinger for suggesting the title of the paper, and Sarbani Basu, Saskia Hekker, Daniel Huber, Benoît Mosser, and Jennifer van Saders for helpful discussion. JMJO acknowledges support from NASA through the NASA Hubble Fellowship grant HST-HF2-51517.001-A, awarded by STScI. STScI is operated by the Association of Universities for Research in Astronomy, Incorporated, under NASA contract NAS5-26555. CJL acknowledges support from NSF grant AST-2205026, and the support of a Gruber Science Fellowship. DS is supported by the Australian Research Council through Discovery Project grant DP190100666. Parts of this work were supported by the Australian Research Council Centre of Excellence for All Sky Astrophysics in 3 Dimensions (ASTRO 3D), through project number CE170100013.

\software{NumPy \citep{numpy}, SciPy stack \citep{scipy}, AstroPy \citep{astropy:2013, astropy:2018, astropy:2022}, Pandas \citep{pandas}, \mesa~\citep{mesa_paper_1, mesa_paper_2, mesa_paper_3, mesa_paper_4, mesa_paper_5}, \gyre~\citep{townsend_gyre_2013}}

\appendix

\section{Analytic approximation for the averaged radial-mode kernel}\label{analytic-approximation-for-the-averaged-radial-mode-kernel}

\label{sec:kernel} We seek an approximation to the expression
\[\begin{aligned}
K &= 2\sum_{n} \sin^2\left[\omega_n t + \delta_0(\omega_n, t)\right]\cdot w(\omega_n)\\
&=1 - \sum_{n} \cos\left(2(\omega_n t + \delta_0(\omega_n, t))\right)\cdot w(\omega_n), 
\end{aligned}\]
where the weights are normalised such that \(\sum_n w(\omega_n) = 1\) --- we recall that \(w\) is described by a Gaussian envelope centered on \(\omega_\text{max} = 2\pi\numax\), with width given by the empirical expression of \citet{mosser_characterisation_2012}. We may restrict our attention to the second term, which we rewrite as
\[\begin{aligned}
\mathrm{Re}&\left[\sum_{n} w(\omega_n) e^{2i (\omega_n t + \delta_0(\omega_n, t))}\right] \\&= \mathrm{Re}\left[e^{2i \delta_0(\omega_\text{max}, t)}\int w(\omega) e^{2i \omega t} \sum_n \delta(\omega - \omega_n)\ \mathrm{d}\omega\right],\label{eq:K_intermediate}
\end{aligned}\]
where we have assumed that \(\delta_0\) does not vary significantly with frequency close to \(\omega_\text{max}\). We recognise this integral as the Fourier transform (from frequency coordinate \(\omega\) to time coordinate \(2t\)) of the product of two different functions of frequency, permitting us to apply the convolution theorem. In detail:

\begin{itemize}
\tightlist
\item
  The first function,
  \[w(\omega) \sim {1 \over \sqrt{2\pi\Sigma^2}} \exp\left[-{(\omega - \omega_\text{max})^2\over 2\Sigma^2}\right],\]
  is a Gaussian of width \(\Sigma = 2\pi\Gamma/8\log 2\). Its Fourier transform is also a Gaussian, modulated by an overall phasor given by the central frequency of this envelope, as
  \[
  \hat{w}(2t) \sim e^{2i \omega_\text{max}t}\exp\left[- 2t^2 \Sigma^2\right].\label{eq:fouriergaussian}
  \]
\item
  The second function is a sum over Dirac delta functions, with \(\omega_n\) approximately uniformly spaced by \(2\pi\Dnu\). We assert that the envelope function \(w\) is narrow enough that the sum may be extended both to arbitrarily high and to negative integer order, without materially changing the value of the integral. A sum over equally spaced Dirac delta functions is a Dirac comb; the Fourier transform of such a comb, with spacing \(\Delta\omega\), is itself a Dirac comb, with spacing \(\Delta (2t) = 2\pi/\Delta\omega = 1/\Dnu \sim 2T\) (given the standard scaling relation for the large separation \(\Dnu \sim 1/2T\)), up to overall constant. As such, we approximate that \cref{eq:fouriergaussian} will have to be convolved against the comb
  \[
  2\pi e^{2i \omega_0 t}\sum_m \delta(2t - 2m T) = 2\pi\sum_m e^{2i \omega_0 m T} \delta(2t - 2m T).\label{eq:comb}
  \]
\end{itemize}

Since the repetition rate of the comb in the position coordinate is the entire acoustic radius of the star, we need only concern ourselves with the \(m=0\) term in \cref{eq:comb} when approximating the average kernel in the inner half of the star. This term is, however, just a Dirac delta function centered at \(t=0\), against which convolution is the identity operation. Thus, \cref{eq:K_intermediate,eq:fouriergaussian} give
\[\begin{aligned}
K &\sim 1 - \mathrm{Re}\left[e^{2i [\omega_\text{max}t + \delta_0(\omega_\text{max}, t)]}\exp(- 2t^2 \Sigma^2)\right], \\
&=1 - \cos\left[2\left(\omega_\text{max}t + \delta_0(\omega_\text{max}, t)\right)\right]\cdot \exp\left(- 2t^2 \Sigma^2\right)\\
&\sim 1 + \cos\left[2\left(\omega_\text{max}t + \arctan\left[\text{softmax}\left({1\over\omega t}, \tan\beta\right)\right]\right)\right]\cdot \exp\left(- 2t^2 \Sigma^2\right).\label{eq:K}
\end{aligned}\]
It is this approximate expression, with the position dependence of \(\delta_0\) specified by \cref{eq:softmax}, which we plot using the dotted curves in \cref{fig:kernelszoom}.

\bibliography{biblio.bib,custom.bib}

\end{document}

%% file: macros.tex
\usepackage{CJKutf8}
\usepackage{newunicodechar}

\newcommand{\Dnu}{\ensuremath{\Delta\nu}}
\newcommand{\numax}{\ensuremath{{\nu_\text{max}}}}

\newcommand\mesa{{\textsc{mesa}}}
\newcommand\gyre{{\textsc{gyre}}}

\newcommand{\chinesename}{{\begin{CJK}{UTF8}{gbsn}(王加冕)\end{CJK}}}

\DeclareRobustCommand{\okina}{%
  \raisebox{\dimexpr\fontcharht\font`A-\height}{%
    \scalebox{0.8}{`}%
  }%
}
\newunicodechar{ʻ}{\okina}

\graphicspath{{./},{./figures/}}
